\documentclass[sigconf]{acmart}

	\author{Seyed-Vahid Sanei-Mehri}
	\affiliation{%
		\institution{Iowa State University}}
	\email{vas@iastate.edu}
	\author{Yu Zhang}
	\affiliation{%
		\institution{Iowa State University}}
	\email{yuz1988@iastate.edu}
	\author{Ahmet Erdem Sar{\i}y\"{u}ce}
	\affiliation{%
		\institution{University at Buffalo}}
	\email{erdem@buffalo.edu}
	\author{Srikanta Tirthapura}
	\affiliation{%
		\institution{Iowa State University}}
	\email{snt@iastate.edu}
	
	\renewcommand{\shortauthors}{Sanei-Mehri, et al.}

\def\BibTeX{{\rm B\kern-.05em{\sc i\kern-.025em b}\kern-.08emT\kern-.1667em\lower.7ex\hbox{E}\kern-.125emX}}

\usepackage{array,epsfig}
\usepackage{bbm}
\usepackage{amsfonts}
\usepackage{amsmath}
\usepackage{amssymb}
\usepackage{amsxtra}
\usepackage{mathrsfs}
\usepackage{datetime}
\usepackage{xcolor} 
\usepackage{mathtools}
\usepackage[noend]{algpseudocode}
\usepackage{tcolorbox}
\usepackage{graphicx}
\usepackage{grffile}
\usepackage[english]{babel}  
\usepackage[utf8x]{inputenc}  
\usepackage{wrapfig}
\usepackage{pifont}
\usepackage{listings}
\usepackage{pgfplots}
\usepackage{footnote}
\usepackage{subfig}
\usepackage[linesnumbered,ruled,boxed,vlined]{algorithm2e}
\SetArgSty{textnormal}
\usepackage{sistyle}
\usepackage{url}

\definecolor{urlclr}{HTML}{0000FF}
\definecolor{bblue}{HTML}{4F81BD}
\definecolor{rred}{HTML}{C0504D}
\definecolor{ggreen}{HTML}{199400}
\definecolor{ppurple}{HTML}{8900ff}
\definecolor{codegreen}{HTML}{094c18}
\definecolor{codegray}{rgb}{9.5,9.5,9.5}
\definecolor{codepurple}{rgb}{0.58,0,0.82}
\definecolor{backcolour}{rgb}{0.95,0.95,0.92}

\makeatletter
\def\thm@space@setup{\thm@preskip=0pt
	\thm@postskip=0pt}
\makeatother
\newtheoremstyle{newstyle}      
{} 
{} 
{\mdseries} 
{} 
{\bfseries} 
{.} 
{ } 
{} 

\theoremstyle{newstyle}
\newtheorem{thm}{Theorem}[section]

\newtheorem{lem}{Lemma}

\makeatletter
\newenvironment{pf}[1][\proofname]{\par
	\pushQED{\qed}%
	\normalfont \topsep0\p@\relax
	\trivlist
	\item[\hskip\labelsep\itshape
	#1\@addpunct{.}]\ignorespaces
}{%
	\popQED\endtrivlist\@endpefalse
}
\makeatother

\newcommand{\E}{\mathbb{E}}

\usepackage[nameinlink,capitalise,noabbrev]{cleveref}
\crefname{lem}{Lemma}{Lemmas}
\crefname{obs}{Observation}{Observations}

\newcommand{\prob}[1]{{\Pr}\left[ #1 \right]}

\newcommand{\wdg}{h}

\newcommand{\ibfly}{\mathbin{\text{\scalebox{1.18}[0.70]{\rotatebox[origin=c]{90}{{\small $\bowtie$}}}}}}
\newcommand{\bfly}{\raisebox{\depth}{$\ibfly$}}

\usepackage{multirow, makecell}
\usepackage{diagbox}
\usepackage{enumitem}

\newcommand{\ignore}[1]{}

\newcommand{\remove}[1]{}

\newcommand{\nbfly}{\xi}

\newcommand{\fleet}{\textsc{Fleet}\xspace}
\newcommand{\ada}{\textsc{Fleet1}\xspace}
 
\newcommand{\adasum}{\textsc{Fleet2}\xspace} 
\newcommand{\iada}{\textsc{Fleet3}\xspace}

\newcommand{\mar}{\textsc{GPS}\xspace} 
\newcommand{\chakra}{\textsc{BC}\xspace} 
\newcommand{\mascot}{\textsc{MASCOT}\xspace}

\newcommand{\bern}{\textsc{Bern}\xspace}

\newcommand{\flip}{\texttt{coin}\xspace} 
\newcommand{\hd}{\texttt{Head}\xspace} 
\newcommand{\reservoir}{\ensuremath{\mathcal{R}}\xspace} 
\newcommand{\peredge}{\textsc{BFC-edge}\xspace} 
\newcommand{\att}{\ensuremath{^{t}}\xspace} 
\newcommand{\attW}{\ensuremath{_W^{t}}\xspace} 

\newcommand{\stream}{\ensuremath{\mathcal{S}}\xspace}

\newcommand{\seqwin}{\textsc{FleetSSW}\xspace} 
\newcommand{\timewin}{\textsc{FleetTSW}\xspace}

\newcommand{\ressize}{\ensuremath{\ensuremath{M}}\xspace}

\newcommand{\adaEx}{\ensuremath{\ensuremath{Y_\ada\att}}\xspace}
 
\newcommand{\adaEst}{\ensuremath{\ensuremath{Y_\ada\att}}\xspace} 
\newcommand{\iadaEst}{\ensuremath{\ensuremath{Y_\iada\att}}\xspace} 

\newcommand{\digg}{\texttt{Digg}\xspace} 
 
\newcommand{\movie}{\texttt{Movie-lens}\xspace} 
 
\newcommand{\yahoo}{\texttt{Yahoo-song}\xspace}
\newcommand{\bag}{\texttt{Bag-pubmed}\xspace}
\newcommand{\frwiki}{\texttt{Edit-frwiki}\xspace}
\newcommand{\enwiki}{\texttt{Edit-enwiki}\xspace}

\newcommand{\vahid}[1]{{\color[HTML]{0000CD}{#1}}} 
\begin{document}

%
\title{FLEET: Butterfly Estimation from a Bipartite Graph Stream}

%
%

%
\begin{abstract}
We consider space-efficient single-pass estimation of the number of butterflies, a fundamental bipartite graph motif, from a massive bipartite graph stream where each edge represents a connection between entities in two different partitions. We present a space lower bound for any streaming algorithm that can estimate the number of butterflies accurately, as well as FLEET, a suite of algorithms for accurately estimating the number of butterflies in the graph stream. Estimates returned by the algorithms come with provable guarantees on the approximation error, and experiments show good tradeoffs between the space used and the accuracy of approximation. We also present space-efficient algorithms for estimating the number of butterflies within a sliding window of the most recent elements in the stream. While there is a significant body of work on counting subgraphs such as triangles in a unipartite graph stream, our work seems to be one of the few to tackle the case of bipartite graph streams.
\end{abstract}


\maketitle

\vspace{1.3ex}
\section{Introduction}
\label{sec:intro}
Enumeration and counting of graph substructures has emerged as a basic tool in understanding complex networks, and has found wide applications in social networks, spam/fraud detection, and link recommendation, and more. Due to the scale of today's datasets, enumeration and counting needs to be performed on very large graphs, with the order of billions of vertices and trillions or larger number of graph substructures. Such large graphs are naturally modeled as graph streams -- the edges of such a graph are not available all at once, but are instead observed as a sequence of updates. 

In this work, we focus on {\bf bipartite graph streams}. A bipartite graph consists of two disjoint node sets $L$ and $R$. Each edge in the graph connects a node in $L$ with a node in $R$. Bipartite graphs are widely used in modeling relationships in the real world. For instance, they can be used to model relationships between authors and papers they have published, where the set of authors form one node partition, papers form the other node partition, and an author has an edge to each paper that she published~\cite{DLK09}. In web search, bipartite graphs have been used in modeling relations between queries and URLs in query logs~\cite{LYLK08} and in matching users to advertisements in computational advertising~\cite{Mehta13,ACMP14}. In computational biology, bipartite graphs are used to model enzyme-reaction links in metabolic pathways and gene-disease associations~\cite{PKP+18}. Other examples include user-product relations, word-document affiliations, and actor-movie networks. Bipartite graphs can be used to represent hypergraphs that capture many-to-many relations among entities, through having the hyperedges in one partition, and the entities in another partition. Bipartite graph streams are natural in the above examples, where new entities may arise in either partition, and new edges are observed as time progresses. The challenge in bipartite graph stream processing is to maintain properties in a time- and space-efficient way as more edges are observed.

While there is a rich literature on subgraph motif counting from unipartite network streams, these methods do not take into account the special structure present in bipartite networks. For instance, the number of triangles (cliques of size 3), a widely studied metric for unipartite graph streams~\cite{Ahmed17, BBC+10, Chen17, Han17, Jha15, Jowhari05, Kutzkov14, Lim15, Pavan13, PTT13a,Shin18, triest, BKS02, Kolountzakis10,Alon97,TT19}, is not a useful metric for bipartite networks, since a bipartite network is triangle-free. Instead, the most basic motif which models cohesion in a bipartite network is the $2 \times 2$ biclique, known as a butterfly~\cite{Aksoy16,Sariyuce18,Vahid18} or a rectangle~\cite{Wang14}. The number of butterflies has been used in defining the clustering coefficient in a bipartite graph~\cite{Robins04,Lind05} and can be considered as playing the same role in bipartite networks as the triangle did in unipartite networks -- a building block for community structure. Though there are some prior works on counting butterflies in a static bipartite graph~\cite{Vahid18,Wang14}, these have not considered bipartite graph streams.

\subsection{Contributions} 
We present FLEET, butterFLy Estimation from a bipartitE graph sTream -- a suite of space-efficient one-pass streaming algorithms for estimating the number of butterflies in a bipartite graph stream. Our algorithms use fixed-size memory that is much smaller than the size of the stream, and continuously maintain an estimate of the number of butterflies as edges arrive in the stream. Our algorithms are simple to implement, backed up by theoretical guarantees, and have good practical performance.

\noindent {\bf -- Space Lower Bound.} We first show a lower bound, proving that any streaming algorithm (whether deterministic or randomized) that can approximately maintain an estimate of the number of butterflies with a bounded relative error in a graph on $n$ vertices must use a memory of size $\Omega(n^2)$ on certain input streams. Note that using $\Omega(n^2)$ memory, it is possible to store the entire graph stream. This shows that in general, it is not possible for a streaming algorithm to maintain an estimate of the number of butterflies using memory sub-linear in the size of the graph. However, the lower bound applies for cases when the number of butterflies in the graph is very small; in particular, the proof depends on distinguishing between two cases, one where there are no butterflies, and another where there is a single butterfly. Real-world bipartite graph streams typically have a large number of butterflies (e.g., \cref{fig:bfly-stream}), and hence, one cannot rule out algorithms that are more space-efficient and return an accurate estimate of the number of butterflies, thus motivating our further work on small-memory algorithms.

\noindent {\bf -- Infinite Window Streaming.} We next present small-memory algorithms for estimating the number of butterflies within an {\em infinite window} consisting of all edges seen so far in the stream. The memory used by our algorithms is no more than a given parameter $M$. We first present an algorithm \ada that is based on adaptive random sampling from the edge stream so that as more edges are seen, the sampling probability decreases so as to fit within available memory. We prove that the estimator returned by \ada is unbiased, and derive concentration bounds showing that the estimator is close to the actual value with a high probability, if the memory used is large enough. We present two enhancements to \ada, leading to algorithms \adasum and \iada which provide better memory-accuracy tradeoffs in practice.

\noindent {\bf -- Sliding Window Streaming.} In stream data mining, the scope of aggregation often needs to be restricted to include edges that have arrived within a recent window. To handle such cases, we present extensions of \fleet to the sliding window model~\cite{BDM02,GT02,BOZ09}. We consider two types of sliding windows. (1)~For a {\em sequence-based} window, defined as the set of $W$ most recent edges in the stream for a window size parameter $W$, we present an algorithm \seqwin  (2)~For a {\em time-based} window, defined as stream elements whose timestamps are greater than $(c-w)$, where $c$ is the current time and $w$ is the window size, we present an algorithm \timewin. Both algorithms use a bounded memory that does not increase with the number of edges in the window. Our algorithm for a time-based window is flexible to receive the window size as a parameter during the query, and does not need to know the window size in advance.

\noindent {\bf -- Experimental Evaluation.} We experimentally evaluate \fleet on real-world graph streams. Results show that our algorithms are effectively able to handle large graph streams. For instance, on the \bag graph with approximately 500M edges and 40T butterflies, our algorithms are able to achieve estimates with an error of less than 1\% using a memory of 600K edges. Our methods present different tradeoffs between memory, estimation accuracy, and runtime, that make them applicable in real-world applications with different requirements, and significantly outperform prior works on subgraph counting from graph streams~\cite{Ahmed17,BC17,Manjunath11}.

\subsection{Related Works}
\label{sec:related}
\textbf{Network motifs.} Network motifs are small subgraphs that are defined on a few nodes and edges. Unlike graph communities or dense subgraphs, whose sizes do not have to be bounded, network motifs are typically subgraphs with less than six nodes. A similar concept is graphlets~\cite{Przulj07}. Network motif detection and counting is now an indispensable tool in network analysis~\cite{Milo02,LY+19}. The distribution of motif counts in a network, as well as the number of motifs that a node takes part in, help characterize the roles of networks and nodes~\cite{Milenkovic08}, an idea that has been used in numerous applications in networking, web and social network analysis, and computational biology~\cite{SSS+15,BBC+10,Faisal14,Wang14,Singh14}.

\noindent \textbf{Butterfly Counting.} There have been relatively few works on counting motifs in a bipartite graph. Wang et al.~\cite{Wang14} presented exact algorithms for butterfly counting in static graphs that outperform generic matrix multiplication based methods~\cite{Alon97}. Sanei-Mehri et al.~\cite{Vahid18} and Zhu et al.~\cite{ZZL18} present exact and randomized algorithms for butterfly counting on static bipartite graphs. \cite{WL+18,SS19} presents parallel algorithms for butterfly counting on static graphs. All these works have considered static graphs and not graph streams, like we do here.

\noindent \textbf{Motif Counting in Graphs.}  There are a number of algorithms known for triangle counting for unipartite streaming graphs, including~\cite{Ahmed17,BBC+10,BKS02, Chen17, Han17, Jha15, Jowhari05, Kutzkov14, Lim15, Pavan13, Shin18, Wang17}. Recent works on counting 4-vertex~\cite{Ahmed15} and 5-vertex subgraphs~\cite{Pinar17} have focused on exact counting and are not designed for streaming graphs. 

Since a butterfly is a 4-cycle, prior works on counting 4-cycles in \emph{any} graph stream~\cite{Bordino08, Buriol07, Manjunath11,Kane12,BC17} can be also applied to a bipartite graph stream, and we compare with such prior work. Note that these algorithms do not use the additional structure in a bipartite graph (absence of edges between vertices in the same partition), and are thus naturally disadvantageous for bipartite streams. Bordino et al.~\cite{Bordino08} present three-pass algorithms for counting 4-cycles, while we focus on single-pass streaming algorithms. Buriol et al.~\cite{Buriol07} consider 3,3-biclique counting from a stream. They assume the {\em incidence stream} model, where edges in the graph stream are presented in a specific order such that all edges incident to a given vertex arrive together, whereas we assume the more general model where edges can arrive in an arbitrary order. Bera and Chakrabarti~\cite{BC17} present algorithms for counting 4-cycles in a graph using two passes through the stream. The work of Ahmed et al.~\cite{Ahmed17} can be specialized to count butterflies in a stream, and we present an experimental comparison with~\cite{Ahmed17,BC17} in Section~\ref{sec:expts}. Other works include Manjunath et al.~\cite{Manjunath11} and Kane et al.~\cite{Kane12} for subgraph counting based on graph sketches.


\noindent \textbf{Lower Bounds for Subgraph Counting.} There are multiple prior works on memory (space) lower bounds for triangle and subgraph counting in general graphs~\cite{BC17,BKS02,CJ17,BOV13}, but not for subgraph counting in bipartite graphs. Since a bipartite graph is more restrictive than a general graph (certain edges are disallowed), lower bounds for unipartite graphs do not directly apply to bipartite graphs. To the best of our knowledge, our work presents the first lower bounds for subgraph counting in bipartite graph streams.

\section{Preliminaries}
\label{sec:prelims}
\begin{figure}
	\includegraphics[width=0.35\textwidth]{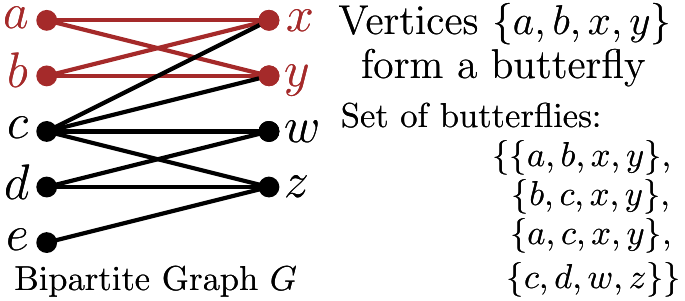}
	\caption{Butterflies in a bipartite graph. Graph $G$ contains four butterflies.\label{fig:bflyexample}}
\end{figure}

We consider simple unweighted and undirected bipartite graphs, without multiple edges between the same pair of vertices. Let $G=(V, E)$ be a bipartite graph with vertices $V$ and edges $E$. The vertex set $V$ of $G$ is partitioned into two disjoint sets $L$ and $R$. The edge set $E \subseteq L \times R$, so each edge $e$ connects one vertex in set $L$ and the other in set $R$.  A \textbf{butterfly} is subgraph on four vertices $\{ a, b, x, y \} \subset V$, where $a, b \in L$ and $x, y \in R$ such that edges $(a, x), (a, y), (b, x)$ and $(b, y)$ exist in the edge set $E$ (see \cref{fig:bflyexample}).

\begin{figure*}
	\includegraphics[width=0.63\textwidth]{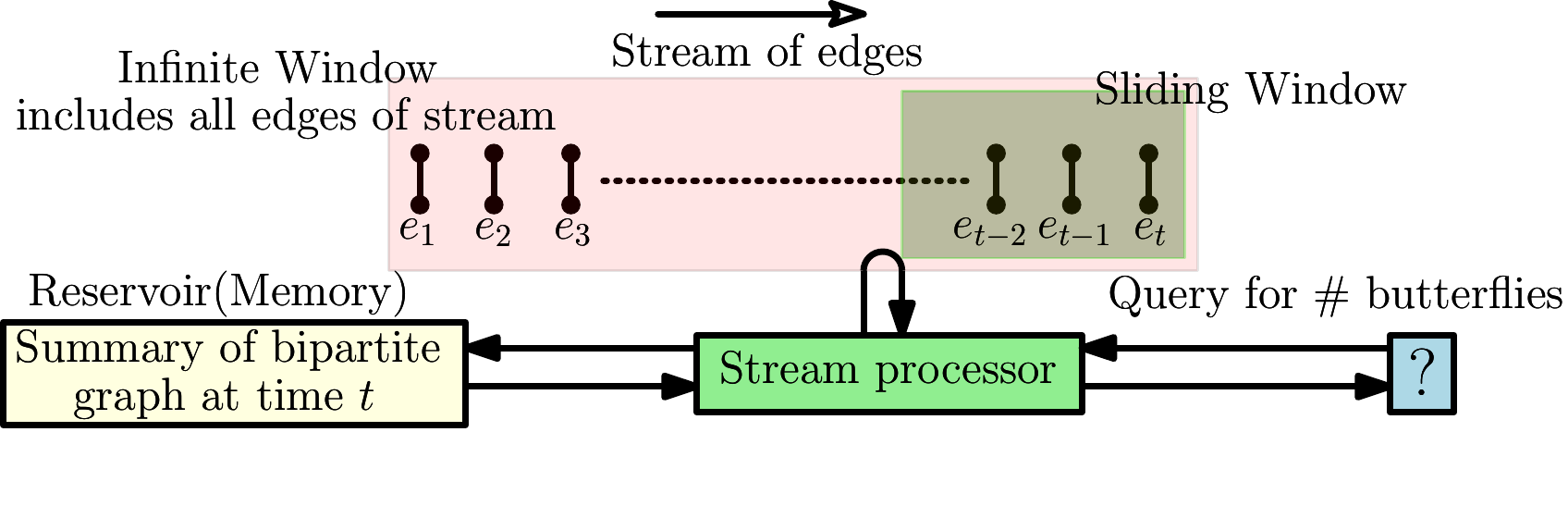}
	\vspace{-5ex}
	\caption{Set up for processing a graph stream.\label{fig:streamdata}}
	\vspace{-0.5em}
\end{figure*}

A \textbf{graph stream} is a sequence of edges $\stream = e_1, e_2, \ldots$ where $e_i$ is the $i$-th edge in the stream. For $t > 0$, let $\stream\att$ denote the first $t$ edges of the stream and let $G\att = (V\att, E\att)$ denote the graph formed by the first $t$ edges, i.e., $G\att = \{ e_1, e_2, \ldots, e_t \}$. 
Let $\bfly\att(G)$ denote the set of all butterflies in the graph $G\att$ and let $\nbfly\att(G) = |\bfly\att(G)|$ denote the number of butterflies in $G\att$. 
When $G$ is clear from the context, we use notations $\bfly\att, \nbfly\att$, and when both $G$ and $t$ are clear from the context, we use $\bfly, \nbfly$. {\cref{fig:streamdata} shows the setup for processing a stream of edges from a bipartite network.}
We consider the following settings.

\noindent \textbf{Infinite Window:} For any $t > 0$ and a stream \stream, the goal is to continuously maintain (an estimate of)  $\nbfly\att$, the number of butterflies in the graph $G\att$, as $t$ changes. 

\noindent \textbf{Sliding Window:} For a window size parameter $W$, a {\em sequence-based sliding window} is defined as the set of $W$ most recent edges. For $t \ge W$, when edge $e_t$ is observed, the sliding window consists of edges $e_{t-W+1}, e_{t-W+2}, \ldots e_{t}$. For $t < W$, the window consists of the entire stream so far. The goal is to continuously maintain an estimate of the number of butterflies in the graph defined by the sliding window. We also consider {\em time-based sliding window},  a generalization of sequence-based window, defined as the set of edges whose timestamps are the most recent. In a time-based window, the window size does not correspond to a specific number of edges, but instead to a range of timestamps.

Our randomized algorithms solely rely on randomness internal to the algorithm, and do not assume that the input is drawn from a specific probability distribution. The input graph stream, including the set of edges and their order of arrivals, could be generated by an adversary. In our space complexity analysis, we assume that a single edge from the graph and an edge timestamp can be stored in a constant number of words. Let $[n]$ denote the set $\{1,2,3,\ldots,n\}$.

\section{Space Lower Bound}
\label{sec:lowerbound}

\newcommand{\anyalgo}{\mathcal{L}}

We show that it is impossible for any streaming algorithm to approximate the number of butterflies to within a small relative error using $o(n^2)$ space. This shows that one cannot expect an algorithm that always returns an accurate estimate using fixed space. In fact, the lower bound shows that essentially, the entire graph needs to be stored (which is possible in $\Theta(n^2)$ bits), if one desires an algorithm that always returns an accurate estimate of the number of butterflies in a bipartite graph stream. Note that our lower bound applies to randomized as well as deterministic algorithms, and for algorithms that return either exact or approximate answers. Note that this lower bound is based on the space complexity of distinguishing between an edge stream that has zero butterflies and one that has at least butterfly.

\begin{thm}
\label{thm:bfly-lb}
For any streaming algorithm $\anyalgo$ that estimates $\nbfly(G)$ for a streaming bipartite graph $G$ on $n$ vertices, there exist input graph streams on which the algorithm uses memory $\Omega(n^2)$ bits.
\end{thm}

The proof uses a reduction from a one-round communication complexity problem, where there are two parties Alice and Bob. Alice gets input $a \in A$ and Bob gets input $b \in B$. It is required to compute a function $g(a,b)$ using only one-way communication from Alice to Bob, while communicating as few bits as possible. $g$ may be computed approximately, and there may be a failure probability that the approximation error is not achieved. Let the one-round communication complexity of function $g: A \times B \rightarrow Z$ with failure probability $\delta$ be denoted by $R_{\delta}^1(g)$.

Consider function $f$ that takes as input $n$ sets $S_1, S_2,\ldots, S_n$ and two integers $i$ and $j$ where each $S_k, 1 \le k \le n$ is a subset of $[n]$ of size $n/10$ and $i,j \in [n]$. The function is defined as: 
\[
f(S_1,S_2,\ldots,S_n, i, j) = 
\begin{cases}
1 & \mbox{if} \qquad j \in S_i \\
0 & \mbox{otherwise} \\
\end{cases}
\]
Inputs $S_1,S_2,\ldots,S_n$ are given to Alice and inputs $i$ and $j$ are given to Bob. Communication is allowed only from Alice to Bob, and Bob has to return the approximate value of the function. We use the following result from~\cite{BKS02}. 

\begin{lem}[Bar-Yossef et al.~\cite{BKS02}]
\label{lem:one-way-cc}
The one round communication complexity of any algorithm for $f$ is lower bounded as follows: for any $0 < \delta < 1/100$, $R_{\delta}^1(f) \ge n^2/40$.
\end{lem}

\begin{pf}[Proof of Theorem~\ref{thm:bfly-lb}]
Suppose there exists a streaming algorithm $\anyalgo$ for estimating the number of butterflies with relative error of $1/2$ with error probability no more than $\delta$, which uses space of $s$ bits. We reduce the one-round distributed computation of $f$ to streaming butterfly counting as follows. 

Given her input $S_1, S_2,\ldots, S_n$, Alice constructs a part of graph $G$ on $4n$ vertices with vertex set $P \cup Q \cup R \cup T$, where $P = \{p_1,p_2,\ldots,p_n\}$, $Q = \{q_1,q_2,\ldots,q_n\}$ and similarly $R$ and $T$. She inserts the following edges into the streaming algorithm in any order. First for each $k=1\ldots n$, edges $(p_k, q_k)$ and $(q_k,r_k)$ are inserted. Next, for each set $S_k, k = 1 \ldots n$, and for each $\ell \in S_k$, an edge is inserted between $t_k$ and $r_\ell$. After Alice is done inserting all these edges into $\anyalgo$, she transmits the contents of the memory of $\anyalgo$ (the entire current state) to Bob, incurring a communication cost of no more than $s$ bits. 

Upon receiving the state of $\anyalgo$ from Alice, Bob continues running $\anyalgo$ by inserting edge $(t_i,p_j)$ into the graph $G$. He then queries $\anyalgo$ for an approximate count of the number of butterflies in $G$. If the answer is non-zero, then he declares that $f(S_1,S_2,\ldots,S_n, i, j) = 1$, and if the answer is zero, then he declares the function to be $0$. Note that if $j \in S_i$, then there is an edge $(t_i,r_j)$ inserted by Alice. If we further consider edges $(t_i, p_j)$ inserted by Bob, and edges $(p_j,q_j)$ and $(q_j,r_j)$ inserted by Alice, we get a single butterfly in $G$. If $j \not\in S_i$ then, it can be verified that there are no butterflies in $G$. Since $\anyalgo$ provides a relative error guarantee, it must return a non-zero estimate if the actual number of butterflies is non-zero and an estimate of zero if the actual number of butterflies is zero. 

We further note that $G$ is a bipartite graph whose vertex set consists of two partitions, $P \cup R$ and $Q \cup T$. It can be verified that there are no edges connecting vertices within a single partition.

Thus, we have reduced the one-round distributed computation of $f$ using $s$ bits of communication to streaming butterfly counting using memory of $s$ bits. Using Lemma~\ref{lem:one-way-cc}, we see that $s \ge n^2/40$, thus completing the lower bound proof.
\end{pf}

\section{Infinite Window Streaming}
\label{sec:algo}
We present streaming algorithms for estimating the number of butterflies over an infinite window, i.e., all edges seen so far. Our randomized algorithms maintain an unbiased estimate of the number of butterflies using a bounded memory of $M$, and provide a trade-off between memory used and the accuracy of the estimate. 

\subsection{Adaptive Sampling: \ada} 
\label{subsec:ada}
We use random sampling from the stream of edges. An initial attempt uses Bernoulli Sampling (\bern) with parameter $p, 0 < p \le 1$. Each arriving edge is sampled into a reservoir with probability $p$. The number of butterflies among the sampled edges is incrementally maintained, and is multiplied by the appropriate normalization factor, to estimate the number of butterflies in the stream. The disadvantage of \bern is that it requires setting parameter $p$ to the ``right value'', which depends on the input itself. If $p$ is too small, the error in estimation can be large, and if $p$ is too large, then the reservoir size can be very large. 

Our first algorithm, \ada, solves this problem by adaptively setting $p$ throughout the computation, so as to keep the memory bounded by $M$. Initially, $p$ is set to $1$, and all edges are sampled into the reservoir. When the size of the reservoir exceeds $M$, $\ada$ sub-samples by retaining each edge in the current reservoir with a probability $\gamma$. Further edges are sampled with probability $\gamma$. When the size of the reservoir again exceeds $M$, the same process is repeated, and further edges are sampled with probability $\gamma^2$, and so on. The size of the reservoir never exceeds $M$ edges, and is at least $\gamma M$, with high probability, except during the initial stages of the stream. \ada also continuously maintains the number of butterflies among sampled edges. When an estimate is desired, the number of butterflies among the sampled edges is returned, after multiplying by the appropriate normalization factor. 

Details are presented in Algorithm~\ref{algo:ada}. Each time the reservoir is sub-sampled, \ada uses an (exact) algorithm to compute $\nbfly(\reservoir)$, the number of butterflies in the reservoir using prior methods designed for a static graph, such as\cite{Vahid18,ZZL18}. \ada also uses an algorithm $\peredge(e, E)$ to count the number of butterflies that contain edge $e$ in the graph induced by edge set $E$. This can be achieved using prior work such as~\cite{Vahid18}. For the purposes of the current discussion, the reader can assume that parameter $\gamma$ is set to $1/2$ -- the main advantages of our algorithm, including bounded sample size and provable accuracy still hold. A modest tradeoff between accuracy and runtime can be obtained by setting $\gamma$ to other values between $1/2$ and $1$, as we discuss further in Section~\ref{sec:gamma}. \cref{lem:exp-ada} shows that \ada maintains an unbiased estimate of the butterfly count after observing each edge. Let \adaEst denote the estimate of $\nbfly\att$ returned by \cref{algo:ada} (\cref{ln:ada-return}) at time $t$. 

\begin{algorithm}[t!]
	\small
	\DontPrintSemicolon
	\KwIn{Edge stream \stream,  max. reservoir size $\ressize$, resampling parameter $\gamma$ (default value of $\gamma = 0.5$)}
	\KwOut{Estimate of $\nbfly\att$, the number of butterflies at $t$} 
	$p \gets 1, \reservoir \gets \emptyset, t \gets 0, \beta \gets 0$\; 
	
	\For{\textbf{each} edge $e$ in \stream}
	{
		$t \gets t + 1$ \;
		\While{$|\reservoir| \ge \ressize$}
		{
			$p \gets \gamma p$\;
			
			\For{\textbf{each} edge $e \in \reservoir$}
			{
				Keep $e$ in \reservoir{} with prob. $\gamma$ and discard with prob. $1-\gamma$
			}
			\color[HTML]{D2368F}{$\beta \gets  p^{-4} \times \nbfly(\reservoir)$}\,\tcp{number of butterflies in $\reservoir$\label{ln:ada-scratch}} 
		}
		\If{\flip$(p)~is~\hd$}
		{
			$\reservoir \gets \reservoir \cup \{e\}$\;
			$\beta \gets \beta +p^{-4}\times \peredge(e, \reservoir)$ \label{ln:normalize-ada}\;
		}
		$\adaEst \gets \beta$   \label{ln:ada-return}\;
	}
	\caption{\ada(\stream,\,$\ressize$): Adaptive sampling}
	\label{algo:ada}
\end{algorithm}

\begin{lem}
	\label{lem:exp-ada}
	$E[\adaEst] = \nbfly\att$
\end{lem}
\begin{pf}
	Suppose the butterflies in $\bfly\att$ are numbered from $1$ to $\nbfly\att$. Let $X_{i}\att (1 \leq i \leq \nbfly\att)$ be a random variable equal to $1$ if all edges of the $i^{\text{th}}$ butterfly appear in \reservoir, and $0$ otherwise. Let $X\att = \sum_{i = 1}^{\nbfly\att}{X_{i}\att}$. From \cref{ln:normalize-ada} of \cref{algo:ada}, we have $\adaEx = X\att \left(p\att\right)^{-4}$. When $t \leq M$, all edges of the stream are sampled, and $X\att = \nbfly\att$.  
	When $t > M$, each edge appears in \reservoir with probability $p\att$. Note that $p\att$ itself is a random variable equal to the sampling probability of an incoming edge at time $t$. To compute $\E[\adaEx]$, we first condition on $p\att$ and then remove the conditioning. 
	
	Since there are four edges in a butterfly, $\E\left[ X_{i}\att \bigm\vert p\att \right] = \left( p\att \right)^4$.
	With linearity of expectation,
	$\E\left[X\att \bigm\vert p\att \right] = \sum_{i = 1}^{\nbfly\att}{\E\left[ X_{i}\att \bigm\vert p\att \right]} = \nbfly\att \left(p\att\right)^{4}$.
	Then, $\E\left[\adaEx \bigm\vert p\att \right] = \E\left[ X\att \left(p\att \right)^{-4} \bigm\vert p\att \right] = \nbfly\att$. By conditional expectation, $\E\left[ \adaEx \right] = \E\left[ \E\left[ \adaEx \bigm\vert p\att \right]\right] = \nbfly\att$. 
\end{pf}

{\bf Concentration analysis of $\adaEst$:} We next show that if the upper bound on the reservoir size ($M$) is large enough, then the estimate $\adaEst$ will be concentrated around its expectation, i.e., have a small relative error, with a high probability. There are a few complexities to deal with here. First, the sampling probability $p\att$ is itself a random variable, since it is the result a random process. We first analyze a simpler algorithm $\bern(p)$ that is based on fixed sampling probability $p$, which we have explained earlier. After $t$ edges, let $Z_p\att$ denote the estimate computed by $\bern(p)$, of $\nbfly(G\att)$, the number of butterflies in $G\att$. Lemma~\ref{lem:ada-bern} shows a concentration bound for $Z_p\att$. The difficulty with analyzing $Z_p\att$ is in handling the dependency between random variables corresponding to different butterflies being sampled into the reservoir. Though different edges are sampled independently by $\bern(p)$, different butterflies are not necessarily independent since butterflies may share edges. We address this with the help of the Hajnal-Szemer\'{e}di theorem, building on ideas from prior works \cite{triest,colorful-triangles,Kolountzakis10} all of who applied this idea in the context of triangle counting.  

\begin{lem}
	\label{lem:ada-bern}
	After $t$ edges, let $h\att$ denote the maximum number of butterflies in $G\att$ that an edge can be part of. For a fixed $p$, and for any $\epsilon \in (0, 1)$, we have
	$
	\prob{ \left| Z_p\att - \nbfly\att \right| \geq \epsilon \nbfly\att} \leq 
	8h\att \cdot exp\left( - \frac{\epsilon^2 \nbfly\att p^4}{12h\att} \right)
	$
\end{lem}

\begin{pf}
	Consider a new graph $H\att = (V_{H\att}, E_{H\att})$ constructed from $G\att$: each vertex $v \in V_{H\att}$ corresponds to a butterfly in $G\att$. For each pair of butterflies $u,v \in V_{H\att}$ that share at least one edge, there is an edge in $E_{H\att}$. Since each edge in $G\att$ can be shared by at most $h\att$ butterflies, the maximum degree of vertex $u \in V_{H\att}$ is $(4h\att-4)$. Using the Hajnal-Szemer\'{e}di theorem~\cite{HS70}, there exists an equitable coloring of $H\att$ with at most $(4h\att-3)$ colors. Let the colors be numbered as $\{1, 2, \ldots, 4h\att-3 \}$. For each butterfly $i \in V_{H\att}$, let random variable $X_i$ be defined as: $X_i = 1$ if all edges of butterfly $i$ are sampled by \bern at time $t$, and $0$ otherwise.
	Let the set of butterflies assigned color $j$ be denoted by $C_j$. For two butterflies $a,b \in C_j$, note that $X_a$ and $X_b$ are independent, since $a$ and $b$ do not share edges, and all their edges are sampled independently. Define $Y_j = \sum_{i \in C_j} X_i$. Note that $Y_j$ is a Binomial random variable since it is the sum of independent 0-1 random variables. We have $\E[Y_j] = |C_j| \cdot p^4$ and $|C_j| > \nbfly\att / 4h\att$, since the coloring of vertices in $V_{H\att}$ is an equitable coloring. Using the Chernoff bound,
	$
	\prob{ \left| Y_j  - |C_j| \cdot p^4 \right| \geq \epsilon \cdot |C_j| \cdot p^4}  
	\leq 	2 \exp \left(- \frac{\epsilon^2}{3} \cdot \frac{\nbfly\att}{4h\att} \cdot p^4 \right) 
	$
	
	Using the union bound,
	\\ 
	{\small 
		$\prob{|Z_p\att - \nbfly\att| \geq \epsilon\nbfly\att} 
		= \prob{\left|\sum_{j=1}^{4h\att - 3} Y_j \cdot p^{-4} - \sum_{j=1}^{4h\att - 3} |C_j|\right| \geq \epsilon \sum_{j=1}^{4h\att - 3} |C_j|}$ \\
		$\leq \sum_{j=1}^{4h\att - 3} \prob{|Y_j \cdot p^{-4} - |C_j|| \geq \epsilon |C_j|)}
		\leq 8h\att \cdot exp\left( - \frac{\epsilon^2 \nbfly\att p^4}{12h\att} \right)$
	}	
\end{pf}

We next derive a concentration result for \ada. \ada essentially returns the estimate due to $\bern(p\att)$ where $p\att$ is itself a random variable. While it is possible to compute the expected value of $p\att$, this cannot be directly plugged into the Lemma~\ref{lem:ada-bern}. Lemma~\ref{lem:ada-concent} shows that for a large enough reservoir size, \ada returns an estimate that is concentrated around its expectation. The proof considers multiple $\bern(p)$ instances, one for each sampling probability, and combines with the event of \ada stopping at one of these levels.
\begin{lem}
	\label{lem:ada-concent}
	Assume $\gamma \leq 0.9$. For any $\epsilon, \delta \in (0, 1)$, when $M$ satisfies
	$M \geq 6t \cdot \sqrt[4]{\frac{12 h\att}{\epsilon^2 \nbfly\att} \cdot \ln \frac{8 h\att (4+\delta)}{\delta}}$
	then $\prob{ \left| \adaEx - \nbfly\att \right| \geq \epsilon \nbfly\att} \leq \delta$.
\end{lem}

\begin{pf}
	Note that the sampling rate of \ada is $\gamma^i$ for $i \ge 0$. 
	We say that \ada is at level $i$ when the sampling rate is $\gamma^i$. 
	For $i \in [0, +\infty)$, define events $S_i$ and $B_i$ as follows. 
	Event $B_i$: Suppose that we execute algorithm $\bern(\gamma^i)$, and we have $\left| Z_{\gamma^i}\att - \nbfly\att \right| \geq \epsilon \nbfly\att$. $S_i$ is the event that \ada stops at level $i$.
	
	Define event $B$: $\left| \adaEx - \nbfly\att \right| \geq \epsilon \nbfly\att$. We decompose the probability of event $B$ in terms of $B_i$ and $S_i$.
	{\small 
		\begin{align*}
			\prob{B} = & \sum_{i=0}^{\infty} \prob{B_i \wedge S_i}
			= \sum_{i=0}^{\ell} \prob{B_i \wedge S_i} + \sum_{i=\ell+1}^{\infty} \prob{B_i \wedge S_i}  \\
			\leq & \sum_{i=0}^{\ell} \prob{B_i} + \sum_{i=\ell+1}^{\infty} \prob{S_i}
		\end{align*}
	}
	where $\ell = \frac{\log (M/6t)}{\log \gamma}$. The sampling probability at level $i$ is $\gamma^{i}$, by Lemma~\ref{lem:ada-bern} we have
	$
	\sum_{i=0}^{\ell} \prob{B_i} = \sum_{i=0}^{\ell} 8h\att \cdot exp\left( - \frac{\epsilon^2 \nbfly\att \gamma^{4i}}{12h\att} \right)
	= \sum_{i=0}^{\ell} \alpha^{\left( \gamma^{-4i} \right)}
	$
	where $\alpha = 8h\att \cdot exp\left( - \frac{\epsilon^2 \nbfly\att \gamma^{4\ell}}{12h\att} \right)$ and $\alpha < 1$.
	
	When $\gamma \leq 0.9$, $\gamma^{-4} > e^{1/e}$, we have $\gamma^{-4i} \geq i$ for any $i \geq 1$. Applying this fact, and using the bound on $M$ we get:
	$
	\sum_{i=0}^{\ell} \alpha^{\left( \gamma^{-4i} \right)}
	\leq \alpha + \sum_{i=1}^{\ell} \alpha^i \leq \frac{2 \alpha}{1 - \alpha} \leq \frac{\delta}{2}
	$

	Let $X_{\ell}$ denote the number of edges in \reservoir when \ada is at level $\ell$. Note that the event \ada stops at level higher than $\ell$ is equivalent to the event that $X_\ell$ is greater than the reservoir size $M$. By $\E[X_\ell] = t \cdot \gamma^{\ell}$ and Chernoff bound, we have
	$
	\sum_{i=\ell+1}^{\infty} \prob{S_i} 
	=\, 
	\prob{X_\ell > M} 
	\le 
	\prob{X_\ell > 6 \cdot \E[X_\ell]}
	\leq \frac{\delta}{2}.
	$ 
	Combining the above bounds, we arrive at $\prob{B} \leq \delta$.
\end{pf}

\subsection{Improved Adaptive: \adasum and \iada}
\label{sec:iada}
We present two algorithms \adasum and \iada which improve upon \ada, providing a better memory-accuracy tradeoff.
\adasum is similar to \ada, but handles sub-sampling differently. Say \ada is at ``level $i$'' when its sampling probability is $\gamma^i$. In \ada, each time the level changes from $i$ to $(i+1)$, edges are discarded according to a random process, and the number of butterflies is recomputed on the new reservoir from scratch (\cref{ln:ada-scratch} of \cref{algo:ada}, shown in pink color). Due to this re-computation, butterflies that were already detected at the higher sampling rate (level $i$) may no longer have all their edges present at the lower sampling rate (level $(i+1)$). In contrast, \adasum does not recompute when the reservoir is sub-sampled. Instead, the current butterfly count at level $i$ is maintained, and as more butterflies are detected at level $i+1$, they are accumulated into this estimate. It can be expected that \adasum obtains a better accuracy than \ada, since it ``catches'' more butterflies than \ada. It is easy to see that the estimation of \adasum is unbiased. In addition to better accuracy, \adasum is also slightly faster than \ada, since it avoids recomputation of butterflies at the sub-sampling step. 

\iada (described in \cref{algo:ada3}) improves accuracy over \ada and \adasum by handling new edges differently. This idea is inspired by Algorithm \mascot of \cite{Lim15}, which used the idea in the context of  counting triangles from a graph stream (the same idea is also used in \cite{triest}). Upon receiving a new edge, the estimate is updated by accounting for butterflies that are created by the new edge (and existing sampled edges), even before deciding whether or not to sample the new edge (see \cref{ln:normalize-iada}). In other words, \iada first updates the estimate and then samples. If the current edge sampling probability is $p$, then the probability of detecting a butterfly involving the new edge increases from $p^4$ (in \ada) to $p^3$ (in \iada). This helps increase the accuracy of butterfly counting while using the same memory. \cref{algo:ada3} has further details. We omit further details and proofs, due to space constraints.

\begin{algorithm}[t!]
	\small
	\DontPrintSemicolon
	\KwIn{Edge stream \stream,  max. reservoir size $\ressize$, resampling parameter $\gamma$ (default value is $\gamma=0.5$)} 
	\KwOut{Estimate of $\nbfly\att$, the number of butterflies at $t$}
	$p \gets 1, \reservoir \gets \emptyset, t \gets 0, \beta \gets 0$ \label{ln:normalize-iada}\; 
	
	\For{\textbf{each} edge $e$ in \stream}
	{
		$t \gets t + 1$ \;
		$\beta \gets \beta +p^{-3}\times \peredge(e, \reservoir)$\;
		\While{$|\reservoir| \ge \ressize$}
		{
			$p \gets \gamma p$\;
			\For{\textbf{each} edge $e \in \reservoir$}
			{
				Keep $e$ in \reservoir{} with prob. $\gamma$ and discard with prob. $1-\gamma$
			}
		}
		\lIf{\flip$(p)~is~\hd$}
		{
			$\reservoir \gets \reservoir \cup \{e\}$
		}
		$\iadaEst \gets \beta$ \label{ln:ada3-return}
	}
	\caption{\iada(\stream,\,$\ressize$): Adaptive sampling}
	\label{algo:ada3}
\end{algorithm}

{\bf Comments:}
\remove{ \vahid{ Here, we briefly summarize and compare the advantages and drawbacks of $\ada$, $\adasum$, and $\iada$. It is worth noticing that the following discussion is backed up by the results of experiments on five real-world bipartite graph streams (See \cref{sec:expts}). As discussed above, $\adasum$ is preferred over $\ada$ as $\adasum$ is a more accurate and slightly faster estimator than $\ada$. $\iada$ is also slower than $\ada$, due to counting the local butterflies of every arriving edge. However, $\iada$ can achieve a better accuracy as more butterflies are captured by the estimator. The experiments in \cref{sec:expts} show that $\iada$ yields $1\%$ error with only $0.5\%$ sample rate in large streams. In addition, it is still fast enough to process tens of thousands of edges per second from a large streams (See \cref{fig:gamma-err-time}). Therefore, one can choose $\iada$ as a highly accurate and reasonably fast estimator. In the case that a higher estimation error is admitted, the faster estimator, i.e. $\adasum$, might be a better option.

}}
Instead of Bernoulli sampling, we could also use reservoir sampling of edges as the basis. We took the route of Bernoulli sampling to simplify the analysis, since it leads to edges being sampled independent of each other (given a sampling probability), unlike reservoir sampling, where the sampling of edges are not independent events. 
Our initial implementations of algorithms based on reservoir sampling without replacement, showed that the accuracy-memory tradeoffs were similar to that of our current algorithms. 

We can also achieve estimates of per-vertex butterfly counts (the number of butterflies that each vertex is a part of) using a similar sampling approach of estimating per-vertex subgraph counts from the reservoir, and maintaining additional state for each vertex. A detailed study of local butterfly counting is a goal for future work.

\section{Sliding Window Streaming} 
\label{sec:slid-wind} 
\newcommand{\seqExact}{\ensuremath{\ensuremath{\nbfly\att_{W}}}\xspace}
\newcommand{\seqEst}{\ensuremath{\ensuremath{Y_{\mathrm{\small sw}}\att}}\xspace} 
\newcommand{\timeExact}{\ensuremath{\ensuremath{\nbfly\att_{W}}}\xspace} 
\newcommand{\timeEst}{\ensuremath{\ensuremath{Y_{\mathrm{\small tw}}\att}}\xspace} 
\newcommand{\argmin}{\ensuremath{argmin}}
In this section, we consider butterfly counting in two types of sliding window models: sequence-based and time-based. We assume the number of edges in a window is (much) greater than the available memory $M$ -- otherwise, the entire window can be stored within memory and an exact algorithm can be applied.

\subsection{Sequence-based Window (\seqwin)} 
\label{subsec:seq-sld}
We first present \seqwin for sequence-based sliding window (\cref{algo:seq-win}), which is based on maintaining a sample of edges from within the sliding window. In the initial stages of observation, all edges fit in memory, and as more edges are observed, we recursively decrease the sampling probability as in \ada, to ensure that the sample fits in memory. However, once the number of edges in a window reaches $W$, it will stay at $W$ henceforth. As a result, when the edge sampling probability $p$ reaches $M/W$, the algorithm does not decrease $p$ any further holds it at $M/W$\footnote{For simplicity of exposition, we assume $M/W$ is a power of $\gamma$.}. The algorithm stores only active edges in the sample, i.e., any item $(e, t')$ such that $t'$ is not within the current window is discarded.  


\begin{algorithm}
	\small
	\DontPrintSemicolon
	\KwIn{\mbox{Edge stream \stream, max. reservoir size $M$, window size $W$ ($\gg M)$}}
	\KwOut{Estimate of $\seqExact$, the number of butterflies in window at $t$}
	$p \gets 1$, $\reservoir \gets \emptyset, t \gets 0$, $\beta \gets 0$\;
	\For{\textbf{each} edge $e$ in $\stream$} {
		$t \gets t + 1$ \;
		\lIf{$p > (M/W)$}{
			Run \ada($\stream$, $M$) and update $p, \beta, \reservoir$
		}
		\Else{
			$p \gets (M/W)$  \;
			\If{\flip$(p)~is~\hd$}
			{
				$\reservoir \gets \reservoir \cup \{e, t\}$, and	$\beta \gets \beta + p^{-4} \times \peredge(e, \reservoir)$  \;
			}
		}
		\vahid{\tcc{Delete expired edges and update estimate}}
		\If{$(e', t') \in \reservoir$ s.t. $t' \leq (t-W)$} {
			$\beta \gets \beta - p^{-4} \times \peredge(e', \reservoir)$, and $\reservoir \gets \reservoir \setminus (e', t')$  \;
		}
		$\seqEst \gets \beta$ \; 
	} 
	\caption{\seqwin(\stream,$\ressize$,$W$): Seq-based SW} 
	\label{algo:seq-win} 
\end{algorithm}

Let $\seqEst$ denote an estimate returned by Algorithm~\ref{algo:seq-win}, of $\seqExact$, the number of butterflies in the window at time $t$.
\begin{lem} 
	\label{lem:seq-win}
	The space of \reservoir in Algorithm~\ref{algo:seq-win} is no greater than $M$ in expectation and $\Pr(|\reservoir| \geq 2M) \leq \left( \frac{e}{4} \right)^M$. 
	$\seqEst$ is an unbiased estimate of $\seqExact$. For parameters $0 < \epsilon < 1$ and $0 < \delta < 1$, when $M \geq 6W \cdot \sqrt[4]{\frac{12 h\att}{\epsilon^2 \seqExact} \cdot \ln \frac{8h\att(4+\delta)}{\delta}}$, 
	then $\prob{ \left| \seqEst - \seqExact \right| \geq \epsilon \seqExact} \leq \delta$.
\end{lem} 

\begin{pf}
	We sketch the proof ideas and omit details, due to space constraints. For the space complexity, note that when $p > M/W$, the algorithm runs \ada and its space is strictly bounded by $M$. Otherwise, $p=M/W$ and the number of edges in the sample is a binomial random variable with parameters $W$ and $M/W$, and the space bounds follow using Chernoff bounds. 
	
	
	At any given time, each edge currently in the window is sampled into \reservoir with probability $p$, and $\E\left[\seqEst\right] = \seqExact \cdot p^4$.
	When $p>(M/W)$, we apply results from \ada for expectation (\cref{lem:exp-ada}) and concentration (\cref{lem:ada-concent}) to show the corresponding properties of $\seqEst$.
	
	When $p=(M/W)$, we rely on an analysis similar to Algorithm~\bern in Lemma~\ref{lem:ada-bern} and derive the concentration result that
	$
	\prob{ \left| \seqEst - \seqExact \right| \geq \epsilon \seqExact} 
	\leq 
	8h\att \cdot exp\left( - \frac{\epsilon^2 \seqExact}{12h\att} \cdot \left(\frac{M}{W}\right)^4 \right)  
	\leq 
	\delta.
	$
\end{pf}

\subsection{Time-based Sliding Window (\timewin)} 
\label{subsec:time-sld}
We next consider the case of a time-based sliding window, where each element has an associated timestamp, and the window at time $t$ consists of all elements with timestamps greater than $(t-W)$, where $W$ is a specified window size. Handling a time-based sliding window is more challenging than a sequence-based window since the number of elements within a time-based window can grow and shrink with time.
The sequence-based window can be seen as a special case of time-based window such that at each time, exactly one edge arrives in the stream. 

\timewin (\cref{algo:time-win}), our algorithm for time-based sliding window, can estimate the number of butterflies when the window size $W$ is provided at query time. We assume an upper bound $n_{max}$ number of edges within a window. Let $T = \lceil 1+ \log_{\gamma} \frac{M}{n_{max}} \rceil$.
\timewin is based on maintaining not only a single sample, as in \seqwin or \ada, but $T$ reservoirs $\reservoir_i, i = 0,1,2,\ldots$, at different sampling rates. Every edge is sampled into $\reservoir_0$. For $i > 0$, each edge that was sampled into $\reservoir_{i-1}$ is sampled into $\reservoir_{i}$ with probability $\gamma$. Each reservoir has a capacity of $M' = M/T$ edges, and contains the most recent edges sampled into the reservoir. Each $\reservoir_i$ is stored as a first-in-first-out queue, so that if a new edge enters when the queue is full, the edge with the earliest timestamp is deleted. 

\begin{algorithm}
	\small
	\DontPrintSemicolon
	\SetKwRepeat{Do}{do}{while} 
	\KwIn{Edge stream \stream, reservoir size $M$, $n_{max} (\gg M)$ (the maximum number of elements within a window)}
	\KwOut{Estimate of $\nbfly(G_W\att)$} 
	$T \gets \lceil 1 + \log_{\gamma} \frac{M}{n_{max}} \rceil$  \;
	\tcp{$d_{\ell}$ is the time of most recent discarded edge in $\reservoir_{\ell}$}
	$\forall i\le T,~\reservoir_i \gets \emptyset, ~d_i \gets 0 $  \;
	\For{\textbf{each} edge $e$ in $\stream$ at time $t$} {
		$\ell \gets 0$ \;
		\Do{\flip$(\gamma)~is~\hd$}
		{ 
			$R_\ell \gets R_\ell \cup \{(e, t)\}$ \; 
			\If{$|R_\ell| > \frac{M}{T}$} {
				$R_\ell \gets R_\ell \setminus \{(e^*, t^*)\}$ s.t. $t^* =\min\{t' |~(e', t') \in R_\ell\}$\;
				$d_{\ell} \gets t^*$  \;
			} 
			$\ell \gets \ell + 1$  \; 
		} 	
	}
	
	\vahid{\tcc{Upon a query window size $W$ at time $c$}}
	$\ell_* \gets \argmin_{\ell \in [T]} \{d_{\ell} \mid d_{\ell} \leq (c-W) \}$  \;
	$\mathcal{A} \gets \{ (e, t') \in \reservoir_{\ell_*} \}$ s.t. $t' > (c-W)$ \tcp{sample of window}
	$\timeEst \gets \gamma^{-4\ell_*} \times \nbfly(\mathcal{A})$  \; 
	\caption{\timewin(\stream, $\ressize$, $n_{max}$): Time-based SW} 
	\label{algo:time-win} 
\end{algorithm}

\begin{lem} 
	$\timeEst$ is an unbiased estimate of $\timeExact$. If \\
	$M \in \Omega\left(\log_{\gamma} \frac{M}{n_{\max}}\sqrt[4]{\frac{n^4_{\max} h\att}{\epsilon^2 \timeExact}  \ln \frac{h\att}{\delta}}\right)$, so $\prob{ | \timeEst - \timeExact| \geq \epsilon  \timeExact} \le \delta$.
\end{lem}

\begin{pf}
	At query time $t$, when presented with a window size $W$, let $G\attW$ denote the graph consisting of all edges that have timestamps in $[t-W+1,t]$.
	For level $\ell \in \{0,1,2,\ldots\}$, let $G\attW(\ell)$ be defined inductively as follows. 
	$G\attW(0) = G\attW$. For $i \ge 0$, $G\attW(i+1)$ is derived from $G\attW(i)$ by choosing each edge in $G\attW(i)$ with probability $\gamma$.
	We note that in Algorithm~\ref{algo:time-win}, $\reservoir_i$ contains the $M/T$ most recent elements from $G\attW(i)$. 
	Further, when a query arrives at time $t$, the algorithm uses $\reservoir_{\ell_*}$ such that $\ell_*$ is the smallest value, where $G\attW(\ell_*)$ is completely contained in $\reservoir_{\ell_*}$.
	
	Let $Z\attW(i)$ denote the estimate of $\nbfly(G\attW)$ derived from $\reservoir_i$. Similar to the proof of \cref{lem:ada-concent}, we define: event $S_i$ is the event that the algorithm chooses $\reservoir_i$ to answer the sliding window query, and $B_i$ is the event that the estimate $Z\attW(i)$ has a relative error that is greater than $\epsilon$. The probability that the algorithm fails to return an estimate that has a relative error within $\epsilon$ is given by $\prob{B} = \sum_{i=0}^\infty \prob{S_i \wedge B_i}$. Using an argument similar to the proof of Lemma~\ref{lem:ada-concent}, we arrive at the result.
\end{pf}

\newcommand{\lgndsz}{0.4}
\newcommand{\tblsz}{0.48}

\section{Experimental Evaluation}
\label{sec:expts}

\begin{figure*}[!t]
	\captionsetup[subfigure]{justification=centering}
	\centering
	\subfloat[\movie]{%
		\includegraphics[width=.2\textwidth] {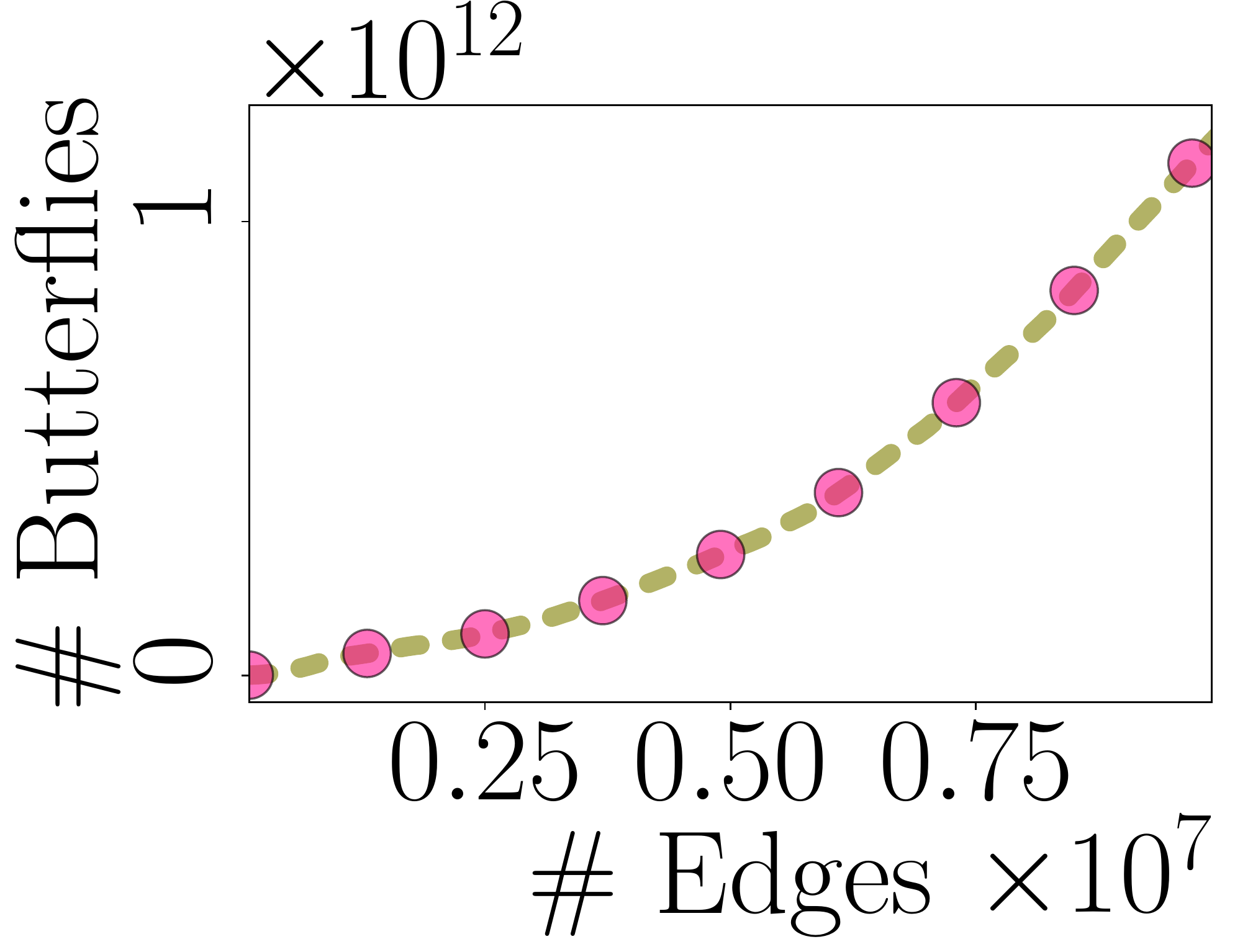}}
	\subfloat[\frwiki]{%
		\includegraphics[width=.2\textwidth] {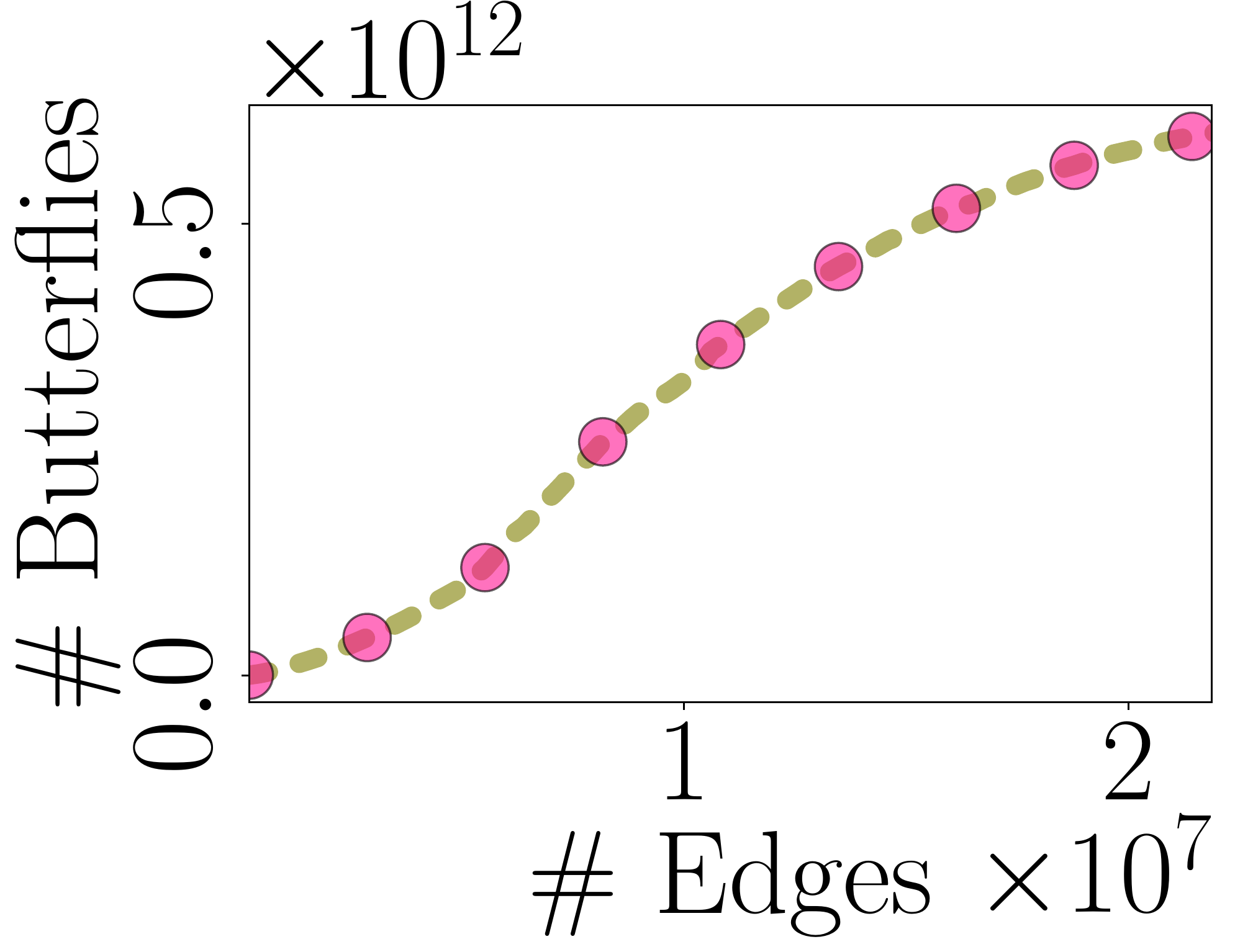}}
	\subfloat[\yahoo]{%
		\includegraphics[width=.2\textwidth] {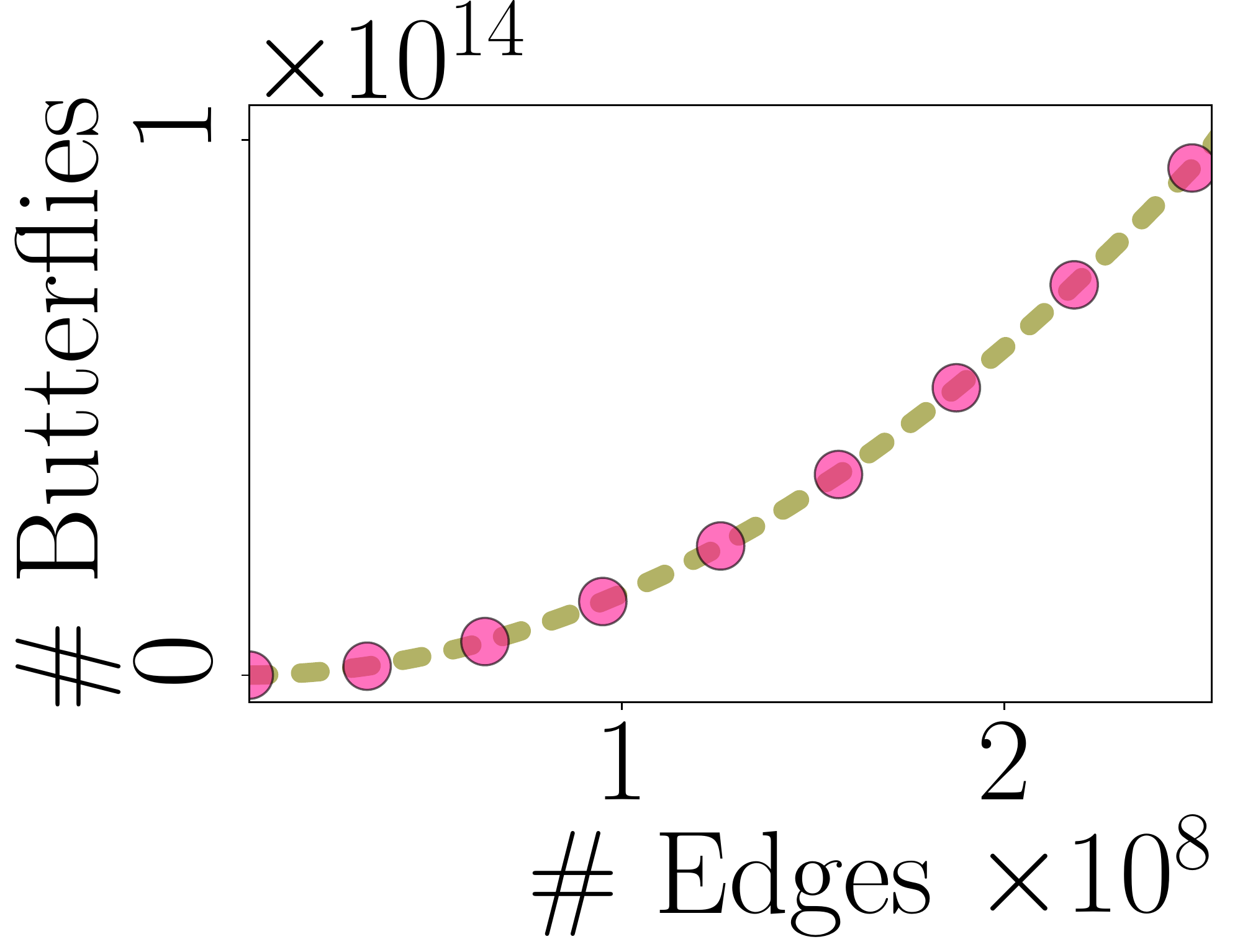}}
	\subfloat[\enwiki]{%
		\includegraphics[width=.2\textwidth]{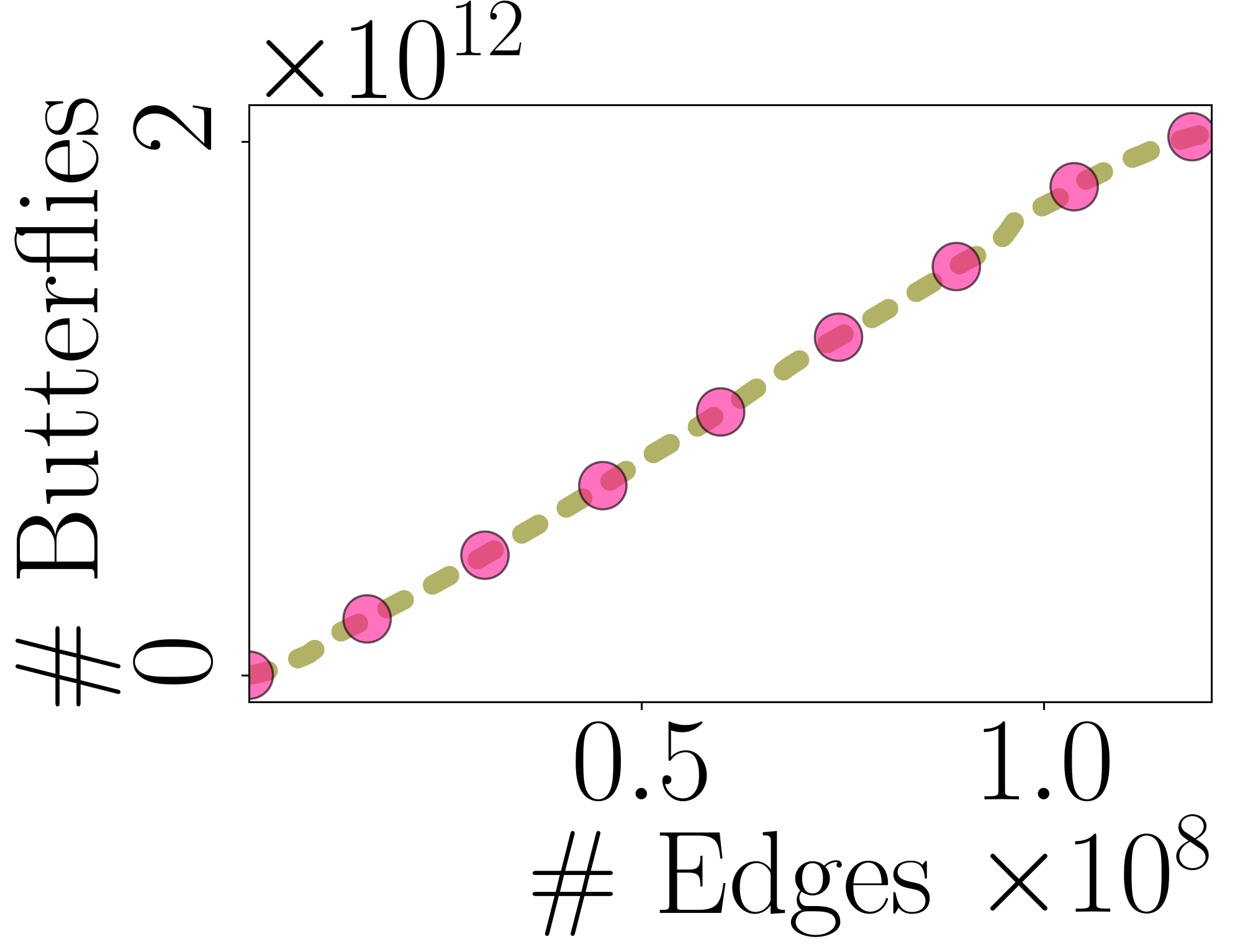}}
	\subfloat[\bag]{%
		\includegraphics[width=.2\textwidth]{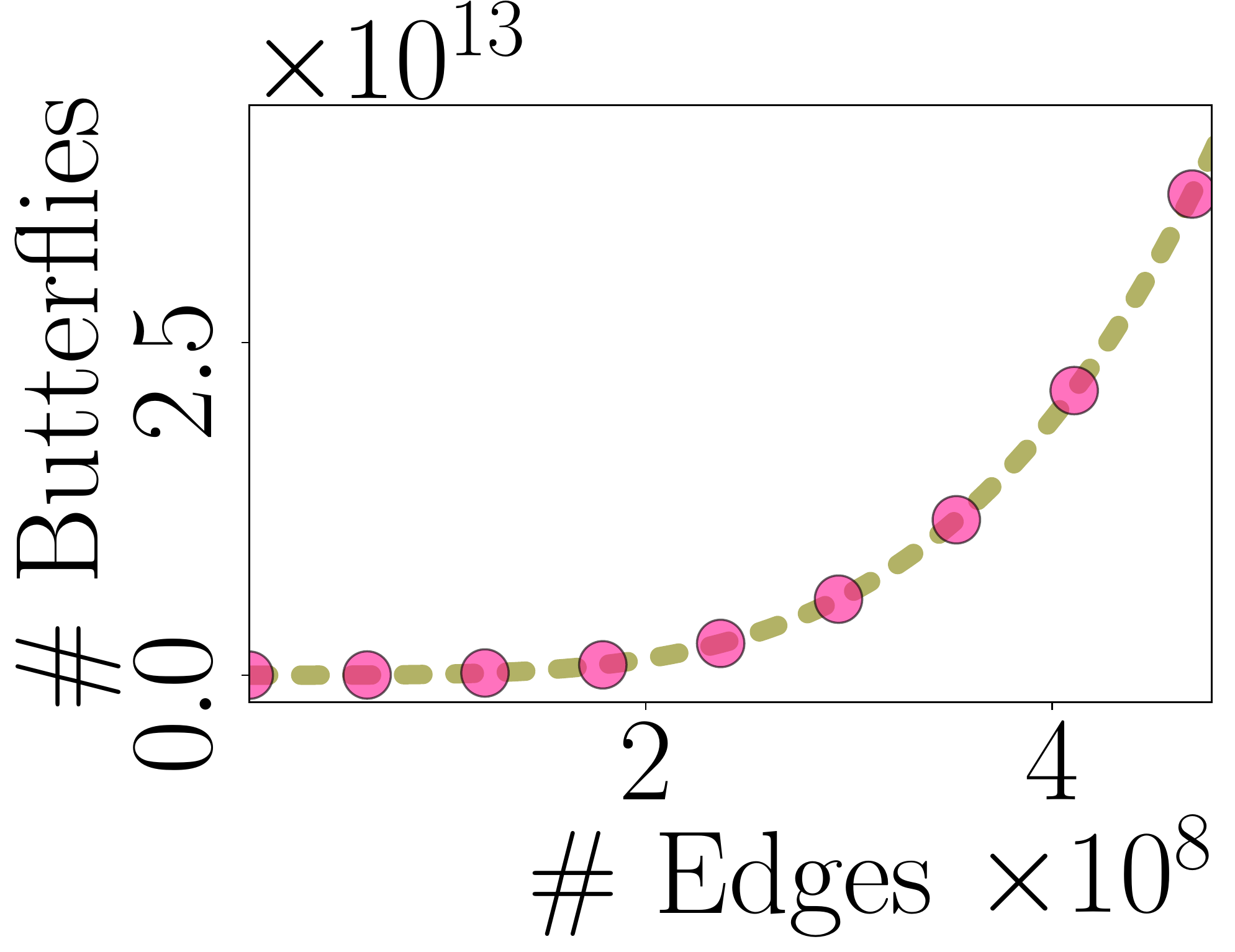}}
		\vspace{-2ex}
	\caption{Number of butterflies as a function of stream size.}
	\label{fig:bfly-stream}
\end{figure*}

\begin{figure*}[!t]
	\vspace{-2ex}
	\captionsetup[subfigure]{justification=centering}
	\centering
	\includegraphics[width=\lgndsz\textwidth] {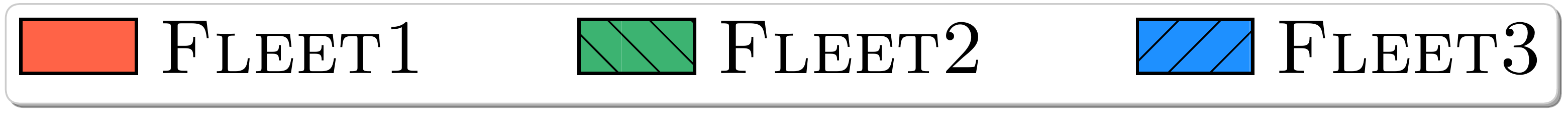}
	\label{fig:legend-err-stream}
	~~~~
	\centering
	\includegraphics[width=0.5\textwidth] {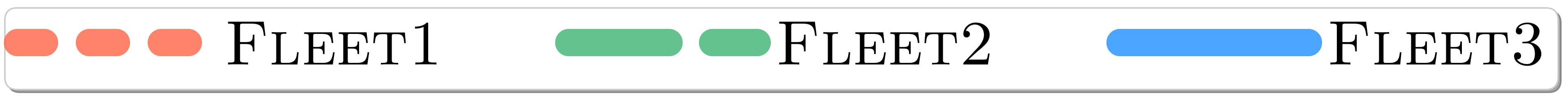}
	\label{fig:legend-err-reservoir}
	\vspace{-2ex}
	\captionsetup[subfigure]{justification=centering}
	\centering
	\subfloat[\movie]{%
		\includegraphics[width=.2\textwidth] {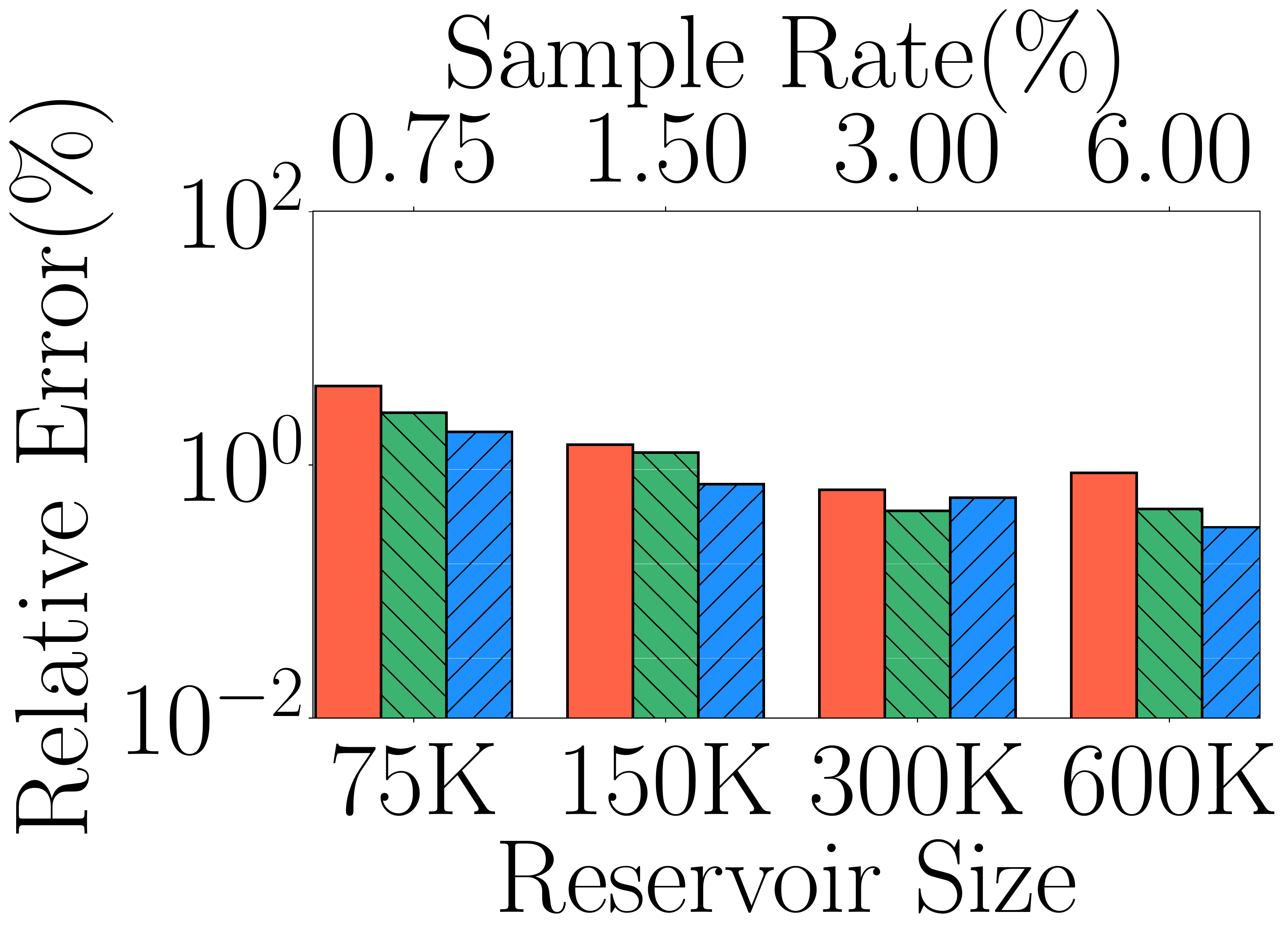}}
	\subfloat[\frwiki]{%
		\includegraphics[width=.2\textwidth]{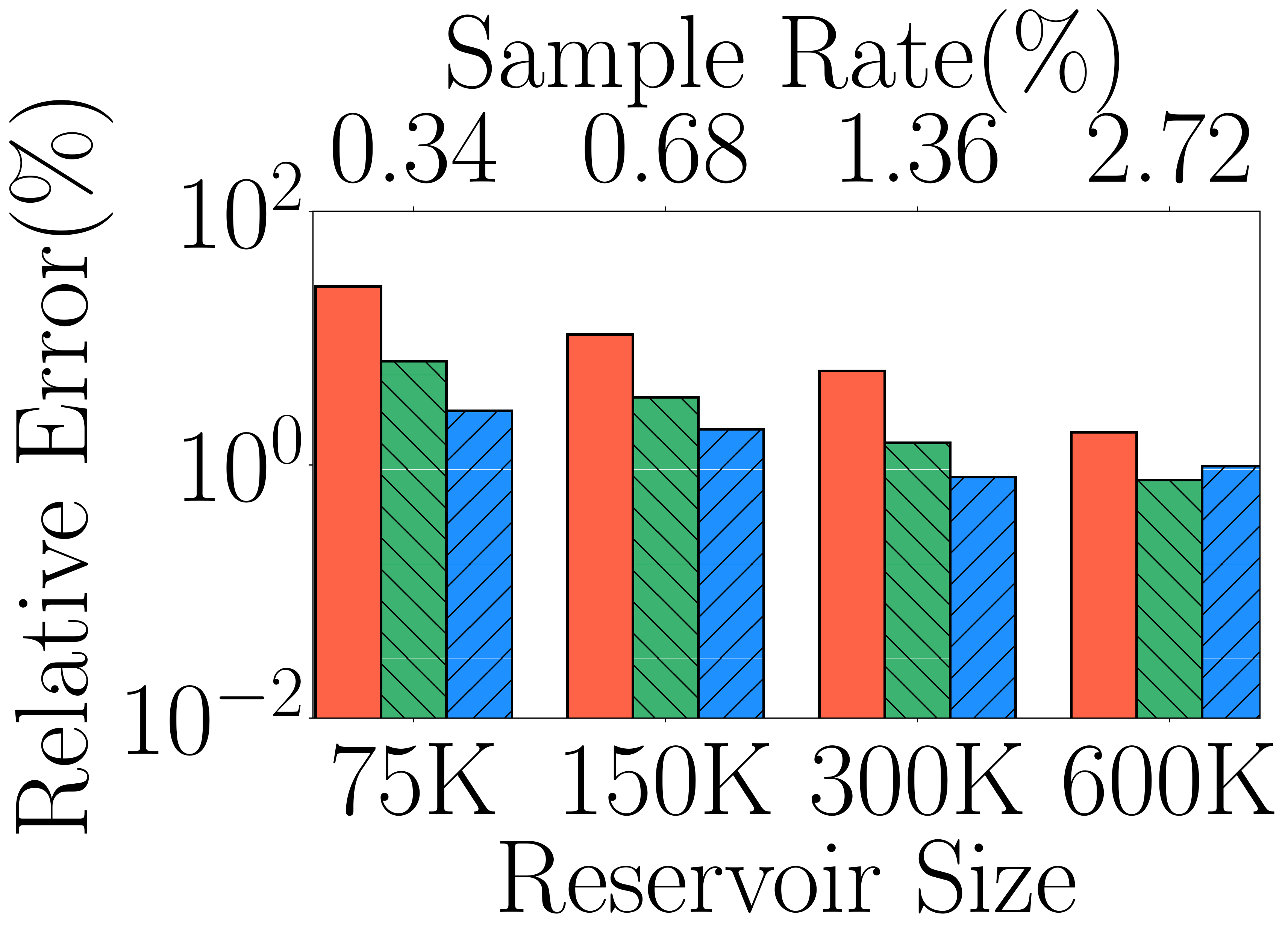}}
	\subfloat[\yahoo]{%
		\includegraphics[width=.2\textwidth]{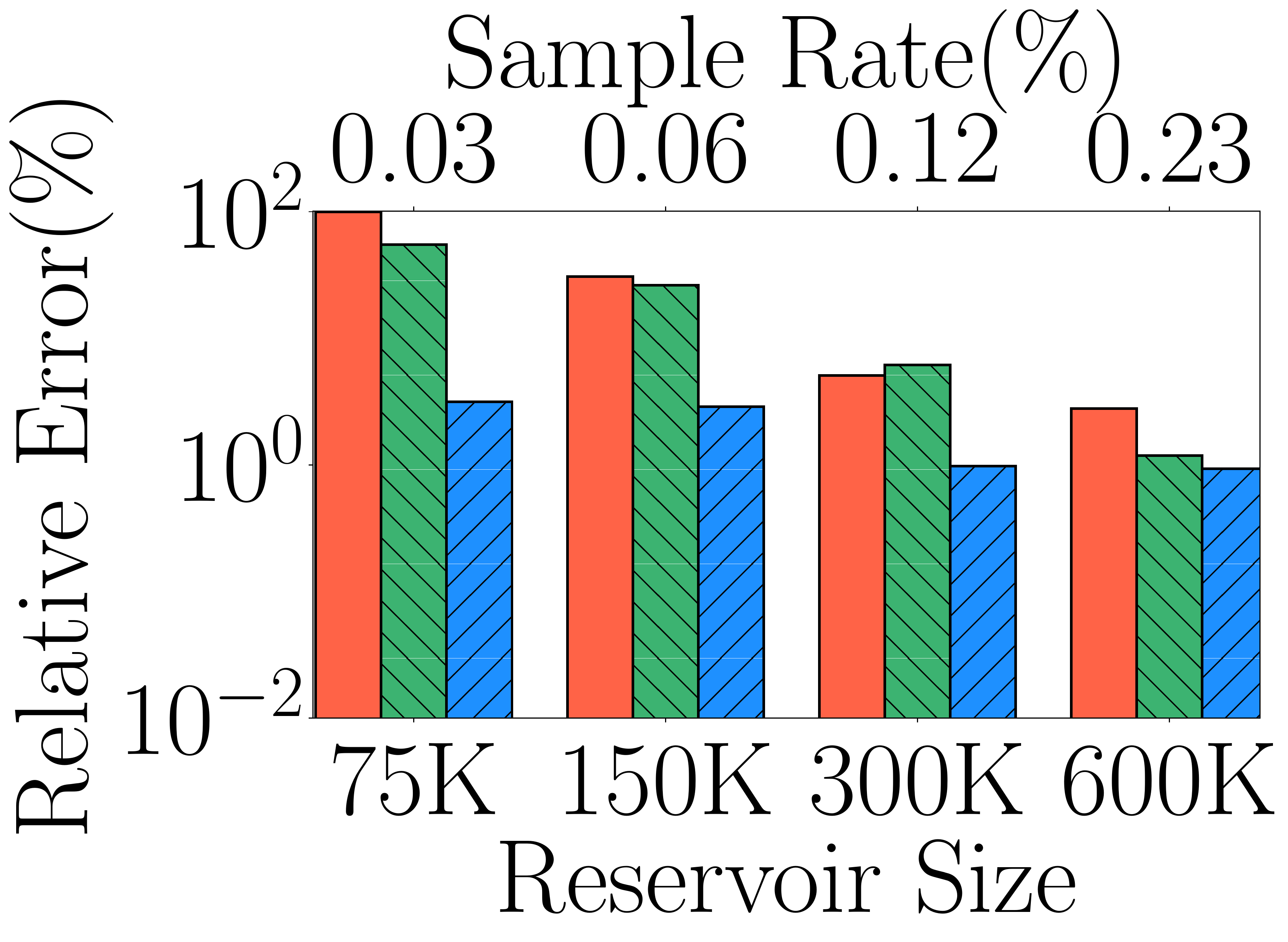}}
	\subfloat[\enwiki]{%
		\includegraphics[width=.2\textwidth] {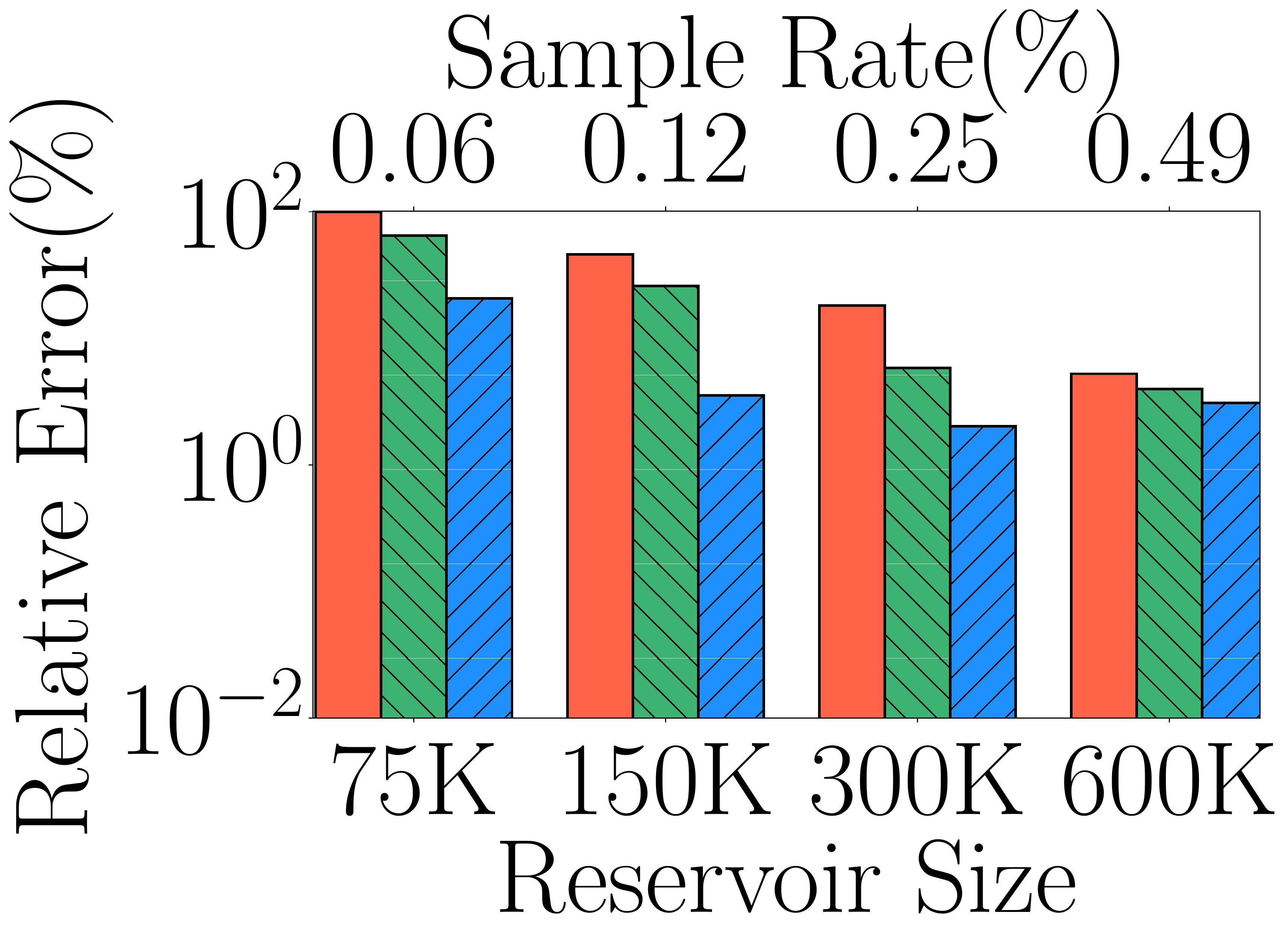}}
	\subfloat[\bag]{%
		\includegraphics[width=.2\textwidth]{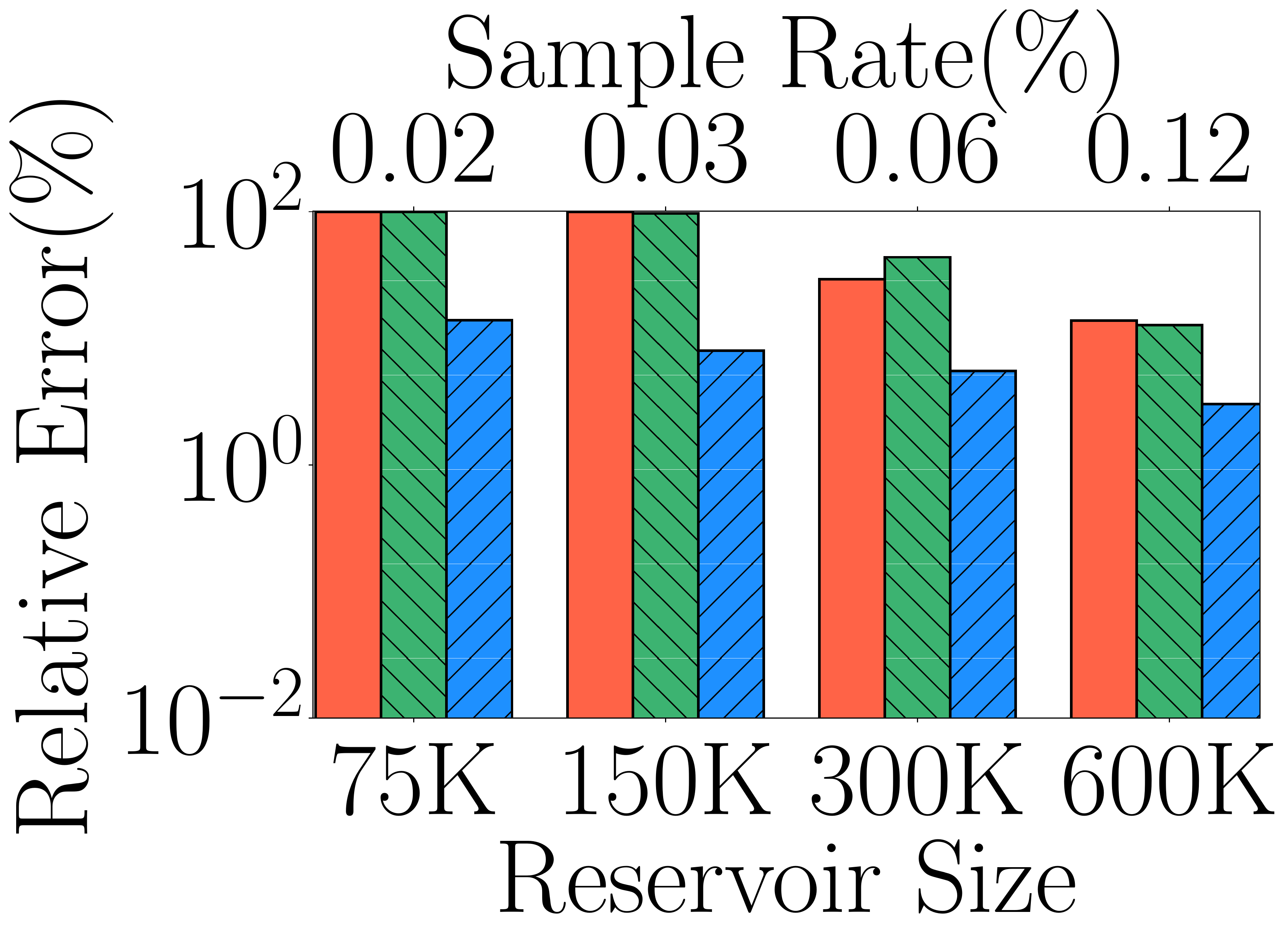}}
		\vspace{-0.5ex}
	\caption{Accuracy of \ada, \adasum, and \iada for $\gamma = 0.5$ versus reservoir size. Bottom x-axis shows the reservoir size and top x-axis shows the sample rate, defined as the ratio of the reservoir size to the stream size.}
	\label{fig:err-reservoir}
\end{figure*}

\begin{figure*}[!t]
	\vspace{-5ex}
	\captionsetup[subfigure]{justification=centering}
	\centering
	\subfloat[\movie]{%
		\includegraphics[width=.2\textwidth] {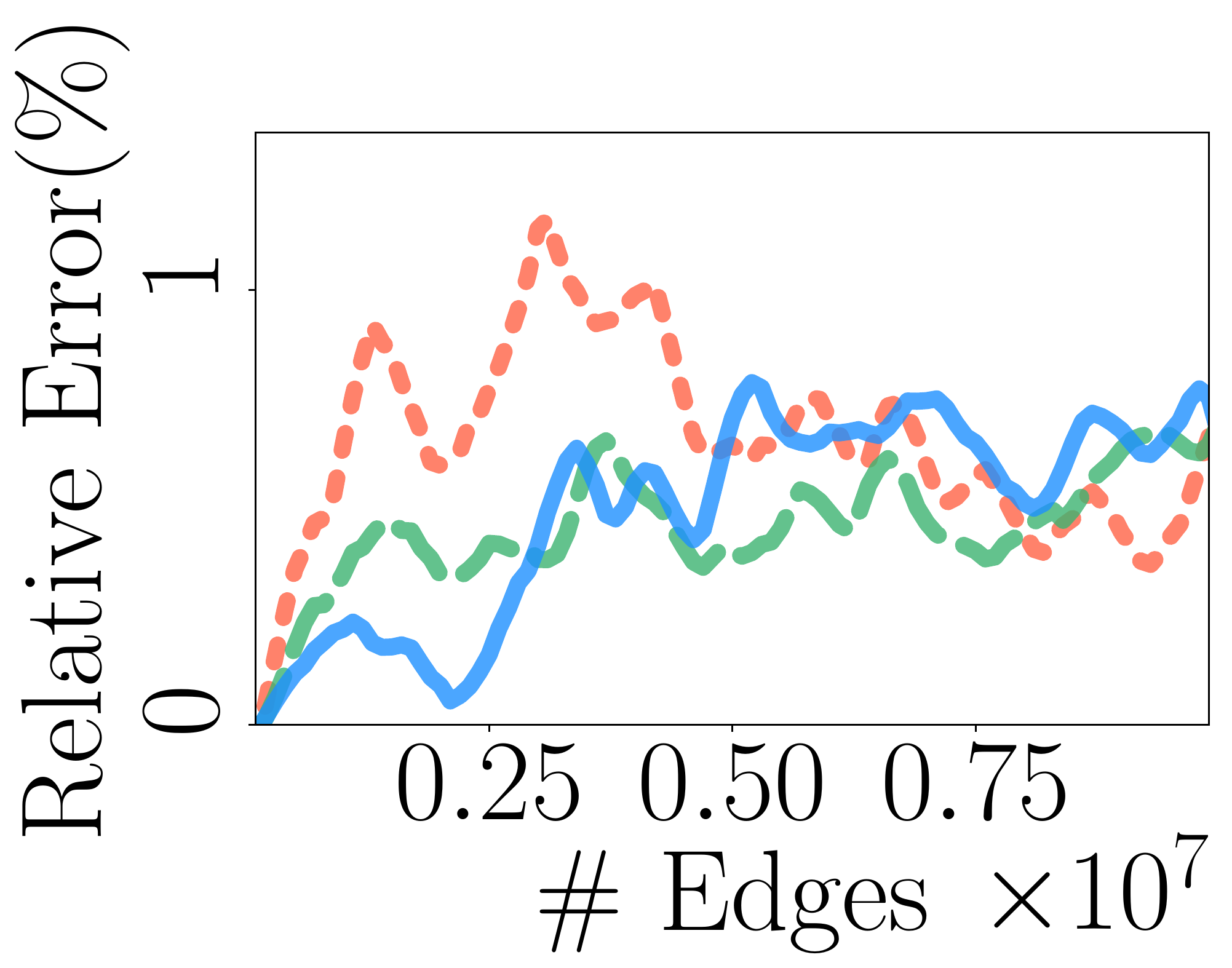}}
	\subfloat[\frwiki]{%
		\includegraphics[width=.2\textwidth]{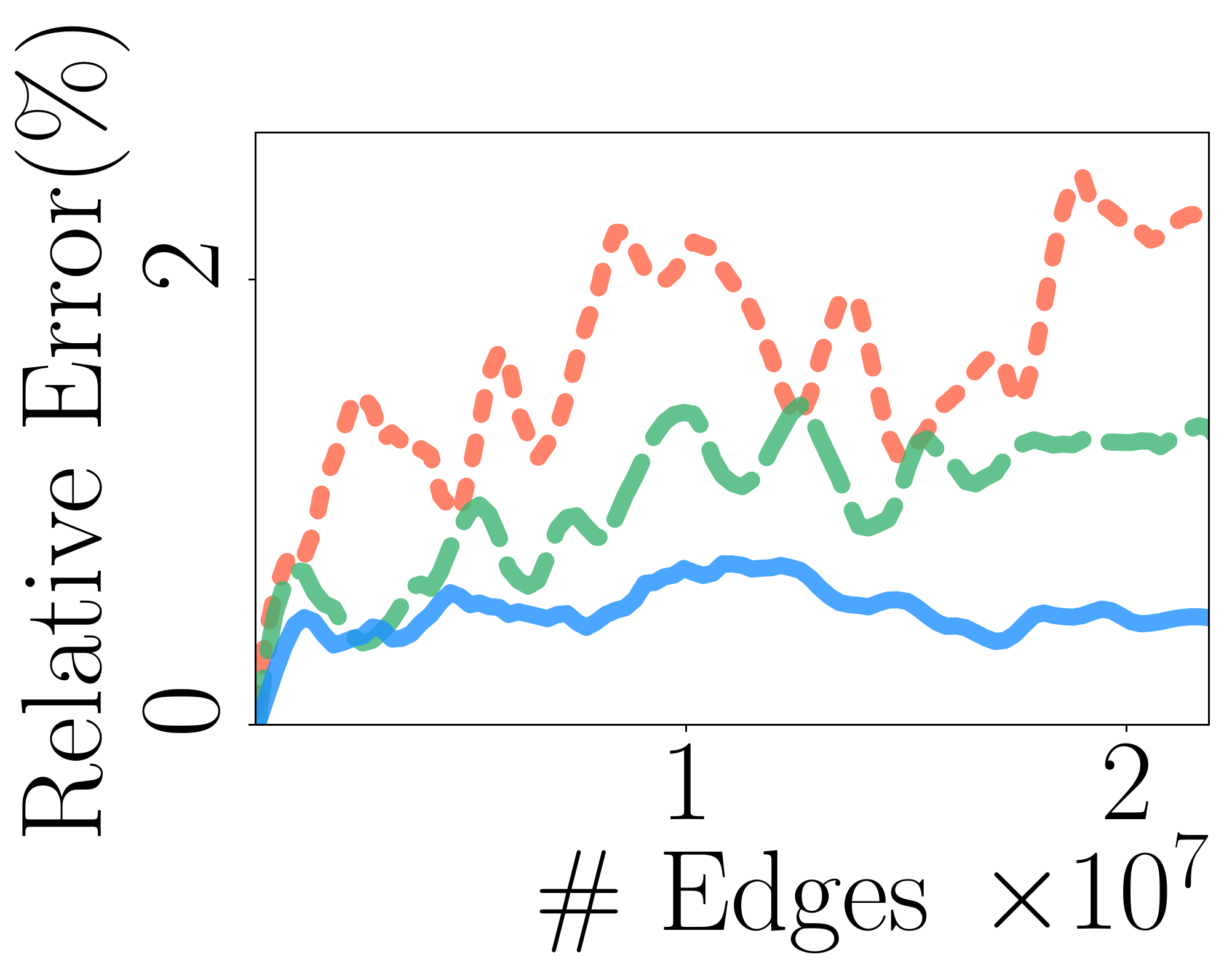}}
	\subfloat[\yahoo]{%
		\includegraphics[width=.2\textwidth]{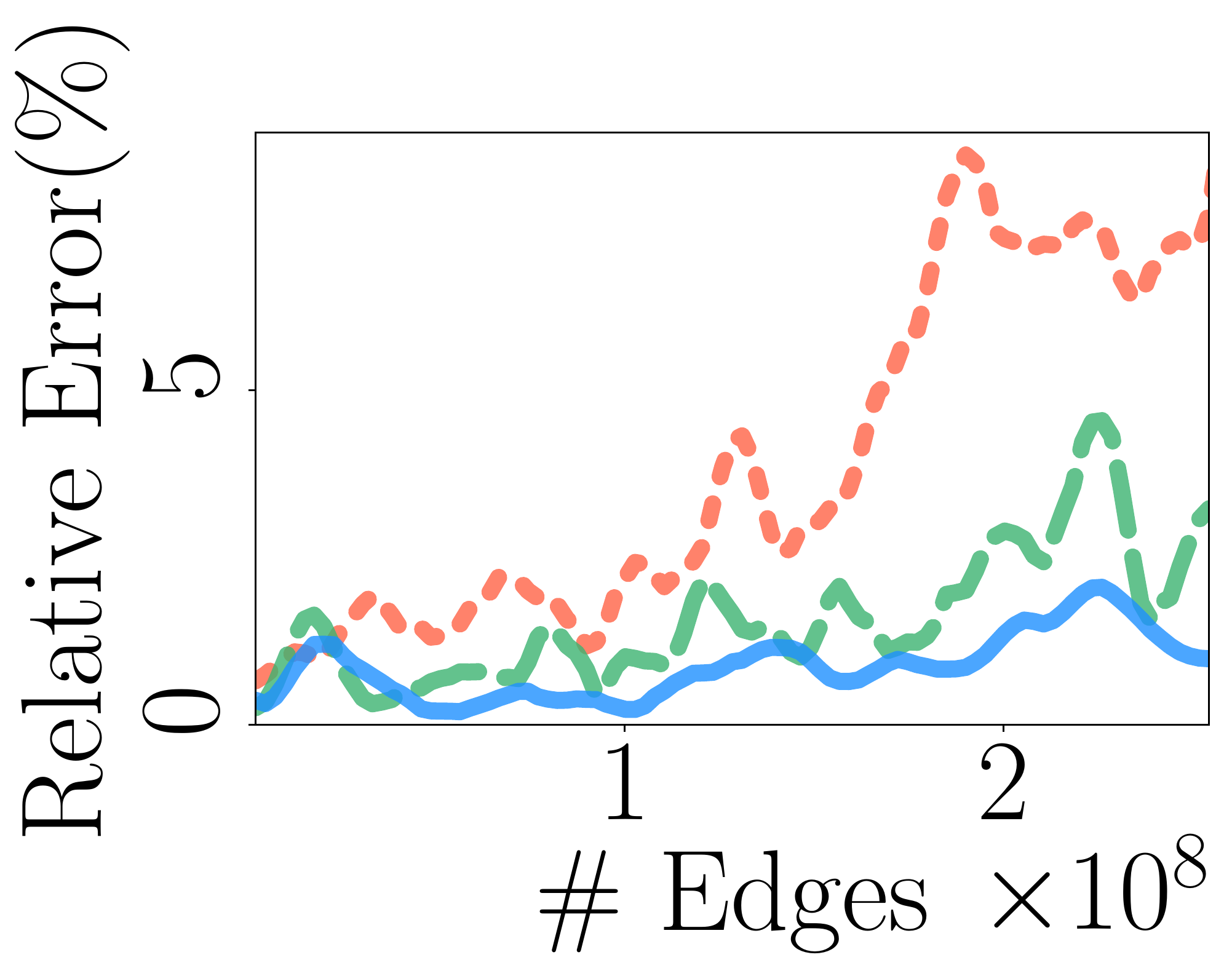}}
	\subfloat[\enwiki]{%
		\includegraphics[width=.2\textwidth] {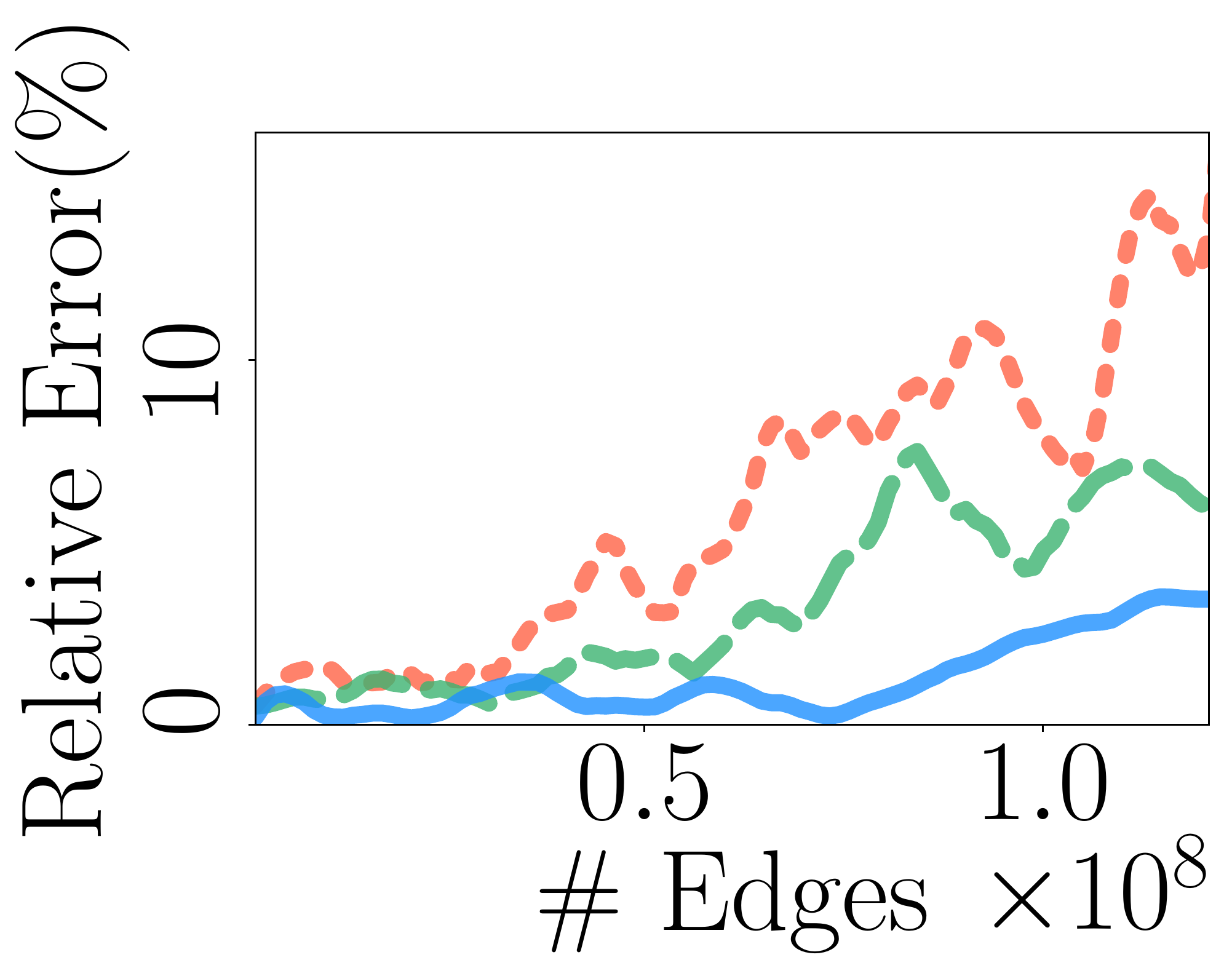}}
	\subfloat[\bag]{%
		\includegraphics[width=.2\textwidth]{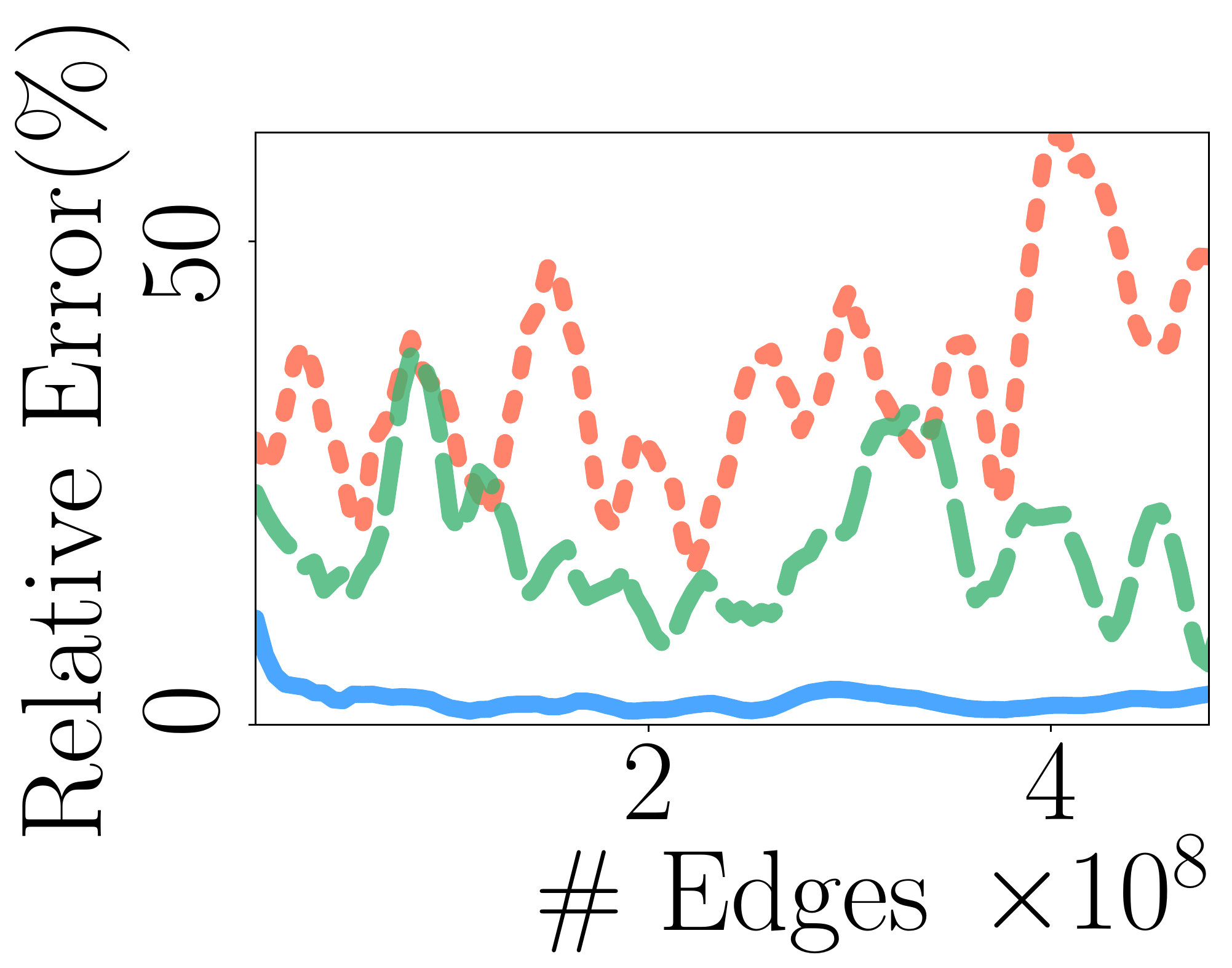}}
				\vspace{-2ex}
	\caption{Accuracy of \ada, \adasum, and \iada at different points in the stream, reservoir size is $\num{300}K$ and $\gamma = 0.9$.}
	\label{fig:err-stream}
\end{figure*}

\begin{figure*}[!t]
	\vspace{-5ex}
	\captionsetup[subfigure]{justification=centering}
	\centering
	\subfloat[\movie]{%
		\includegraphics[width=.2\textwidth] {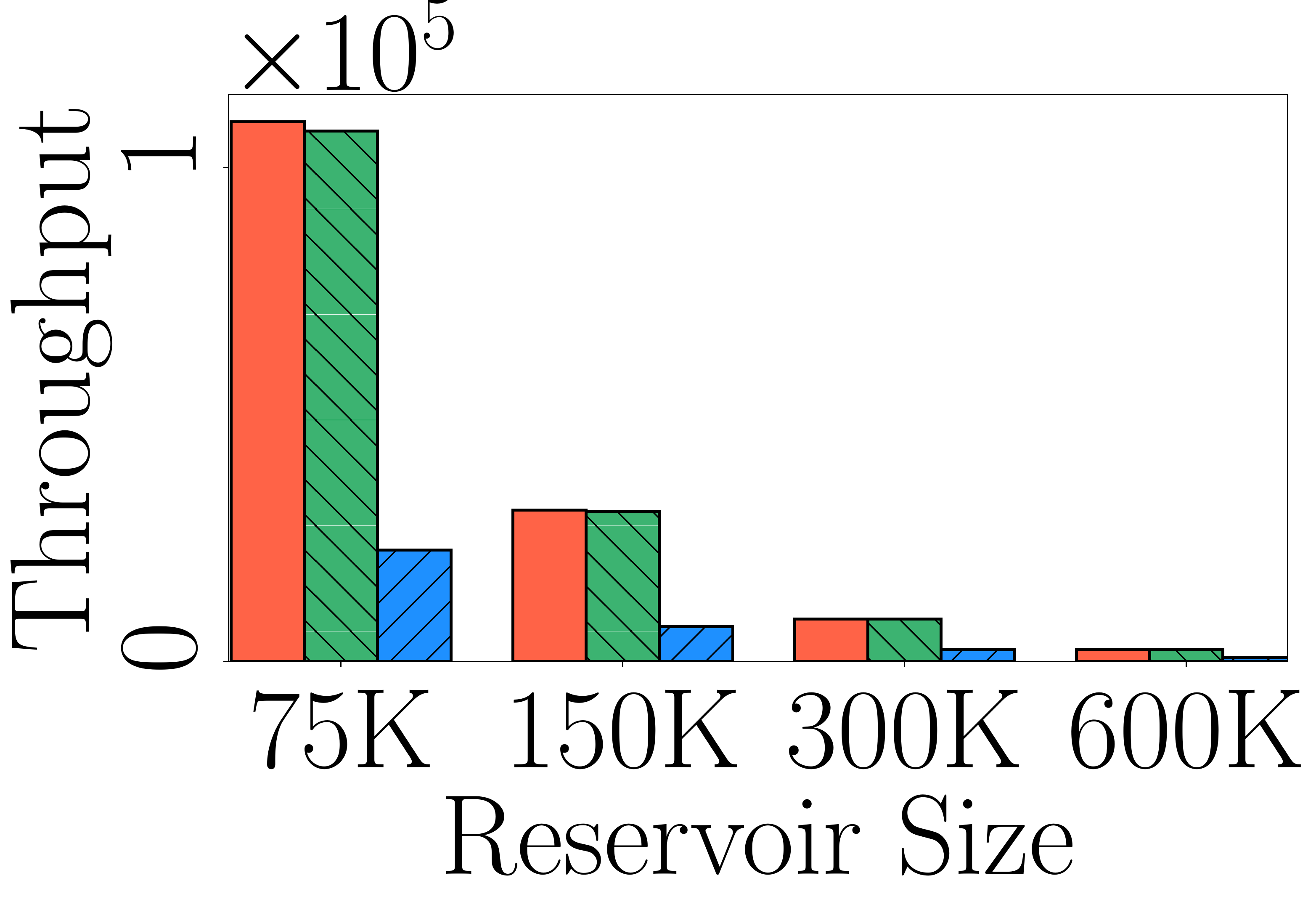}}
	\subfloat[\frwiki]{%
		\includegraphics[width=.2\textwidth]{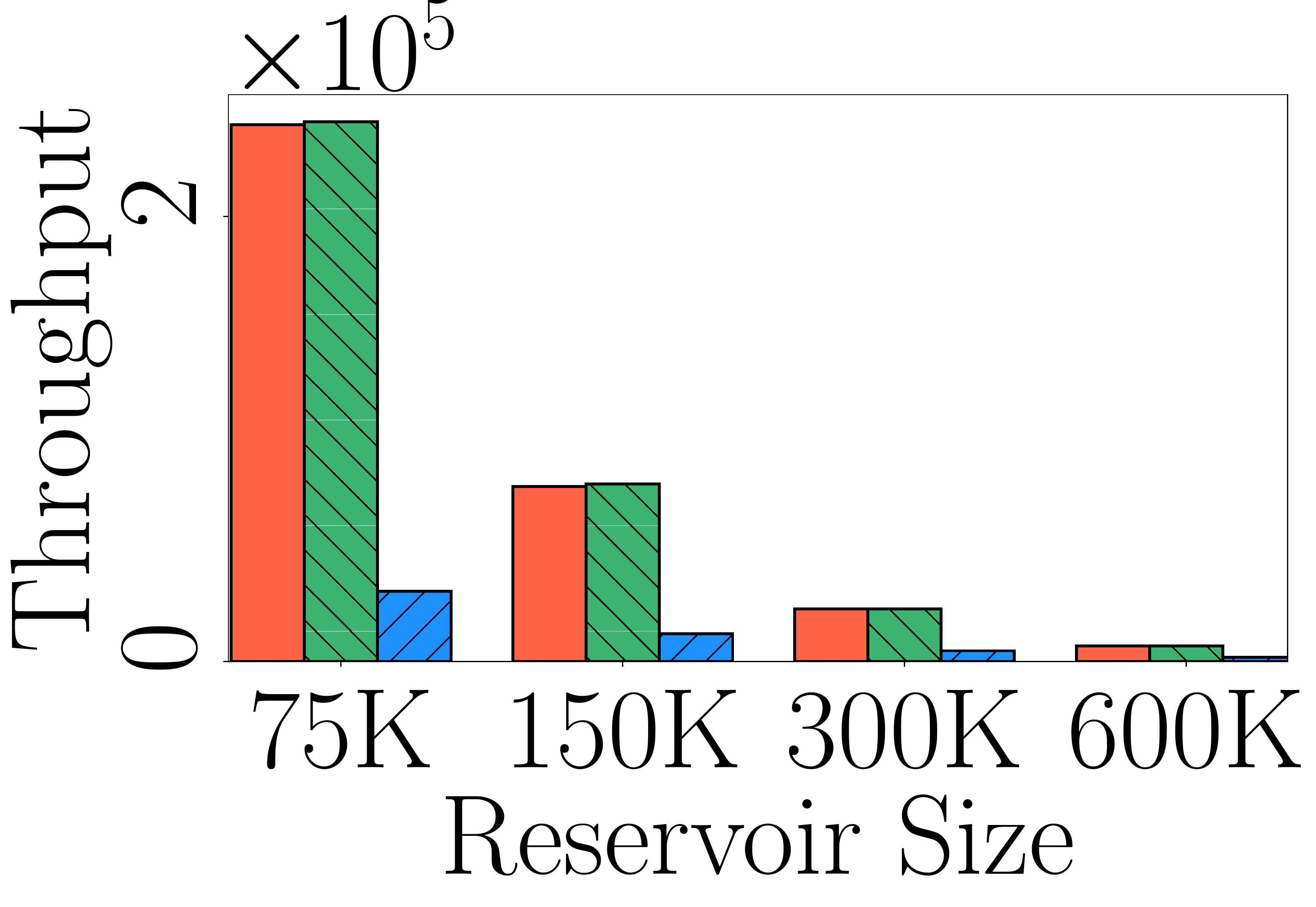}}
	\subfloat[\yahoo]{%
		\includegraphics[width=.2\textwidth]{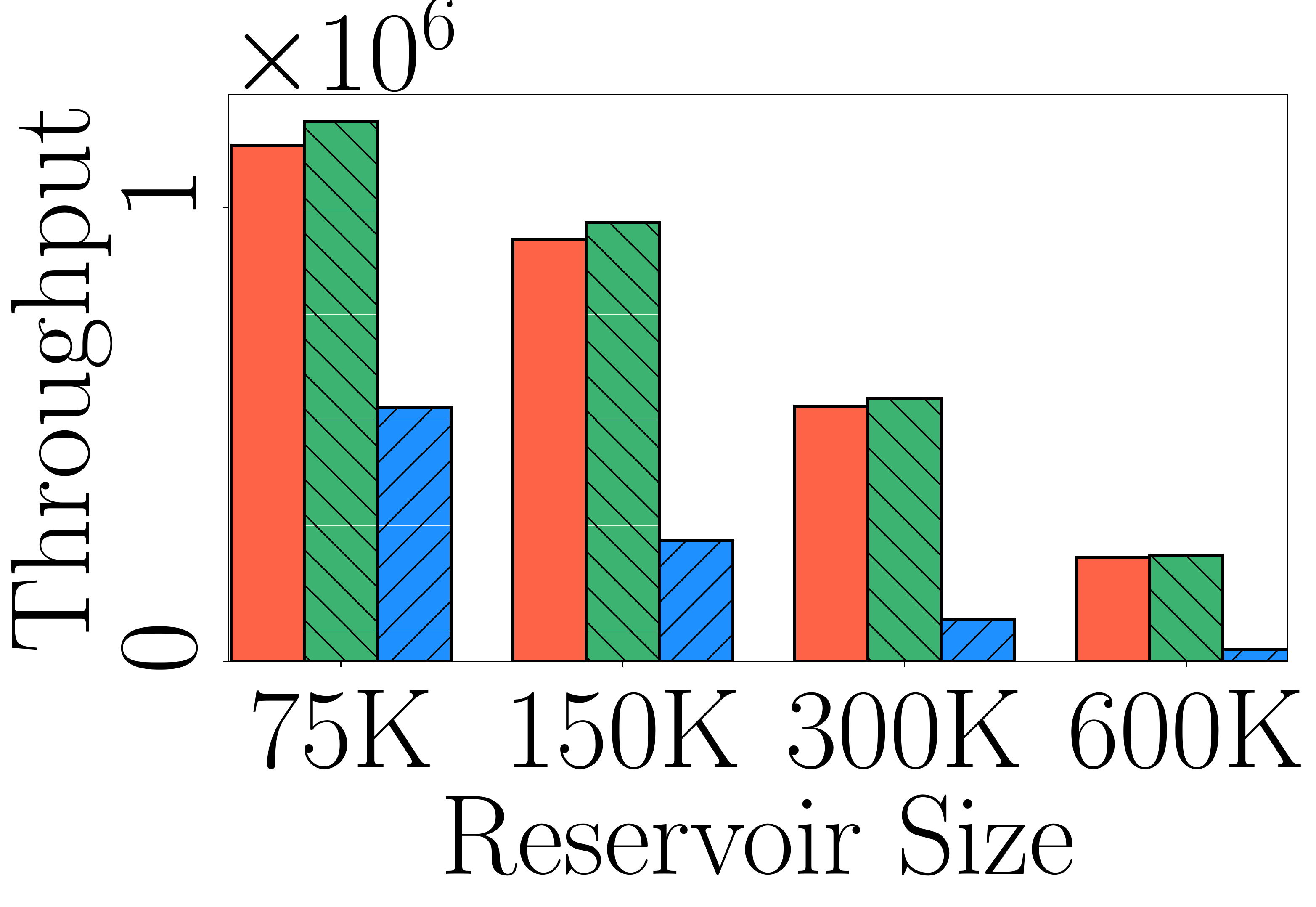}\label{fig:yahoo-throughput-reservoir}}
	\subfloat[\enwiki]{%
		\includegraphics[width=.2\textwidth] {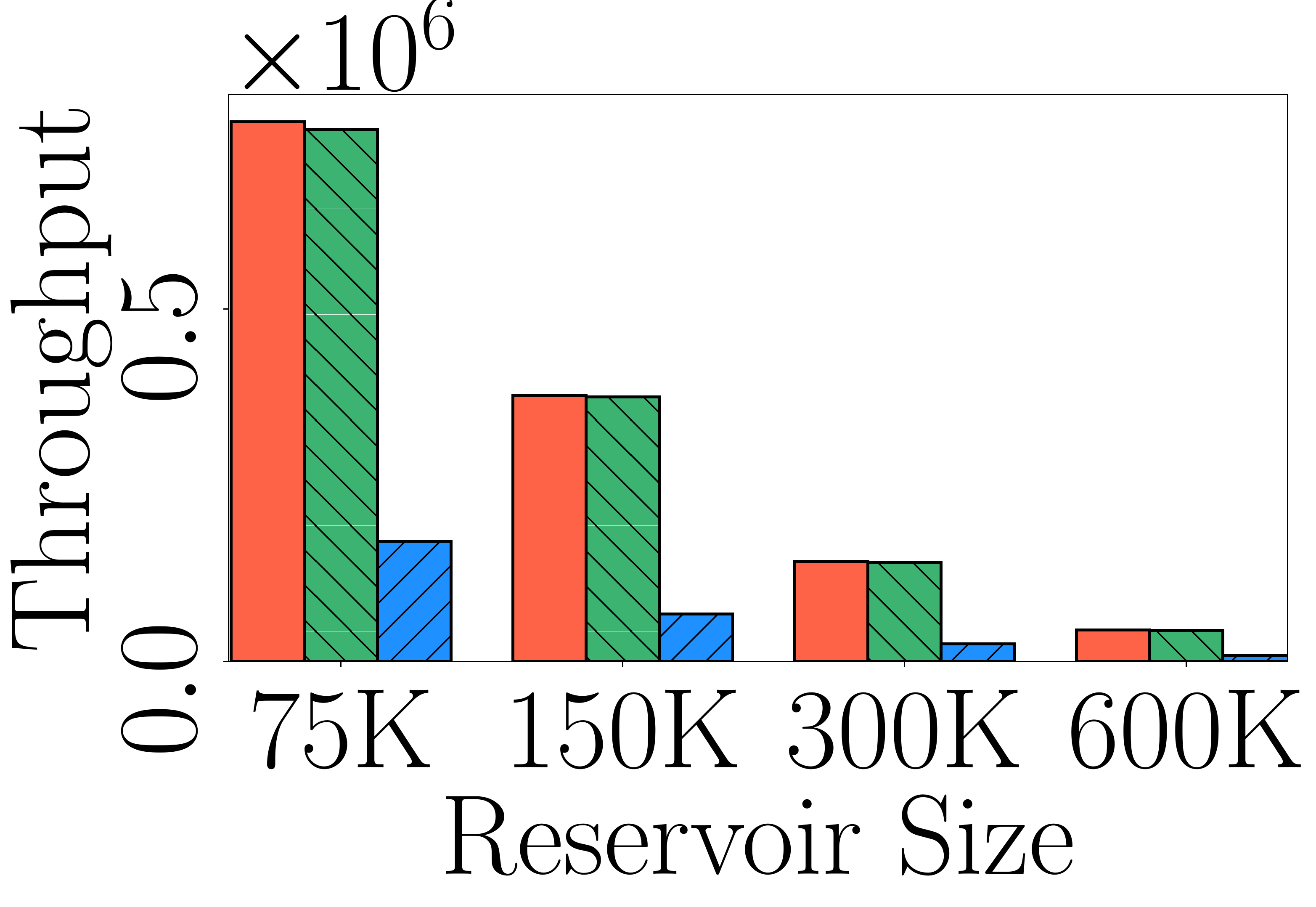}}
	\subfloat[\bag]{%
		\includegraphics[width=.2\textwidth]{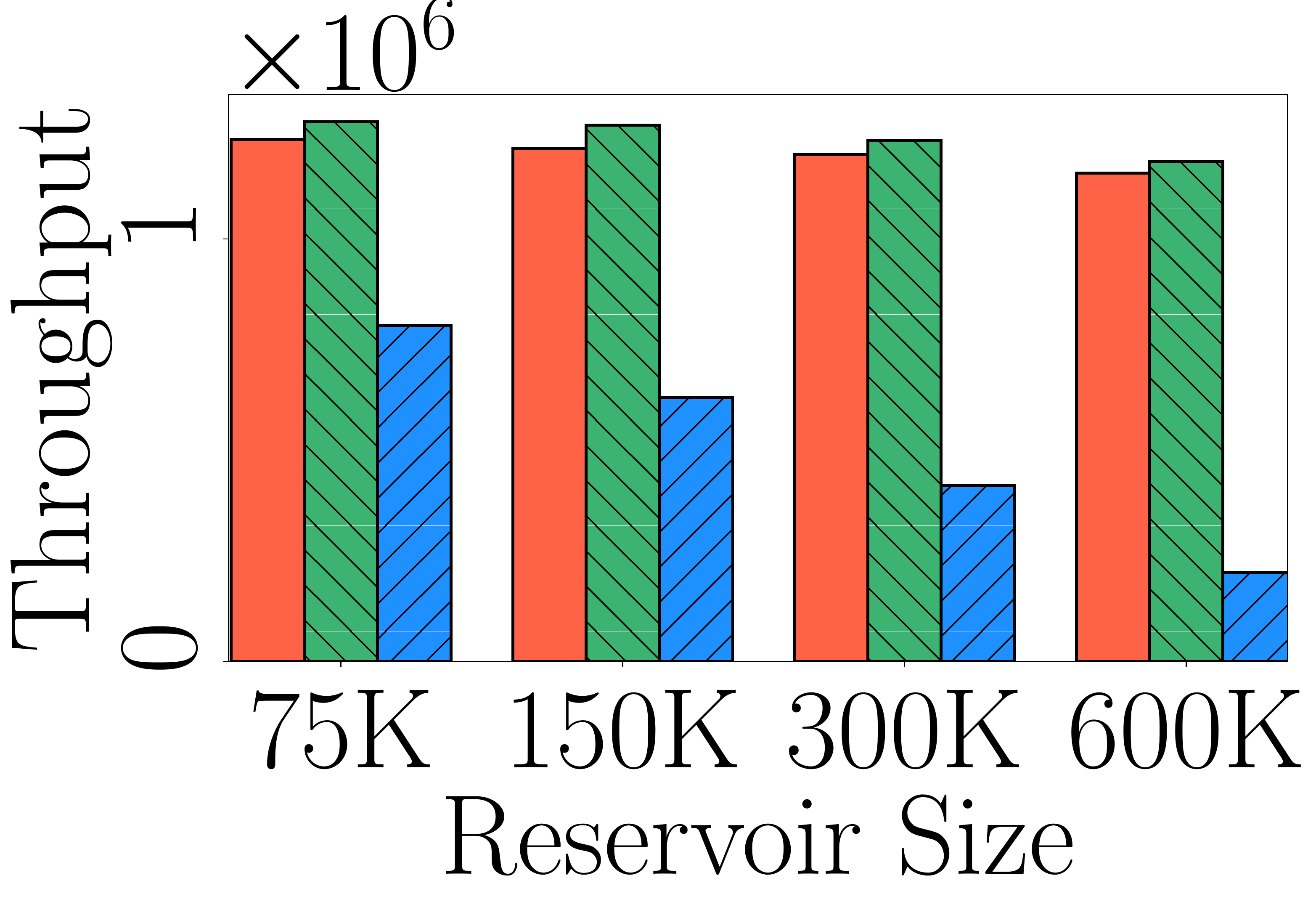}\label{fig:bag-throughput-reservoir}}
	\vspace{-2ex}
	\caption{Throughput of \ada, \adasum, and \iada algorithms as a function of reservoir size where $\gamma = 0.6$.}
	\vspace*{-0.2cm}
	\label{fig:throughput-reservoir}
\end{figure*}
\begin{figure*}[!t]
	\vspace{-5ex}
	\captionsetup[subfigure]{justification=centering}
	\centering
	\subfloat[\yahoo (error vs. $\gamma$)]{%
		\includegraphics[width=.25\textwidth] {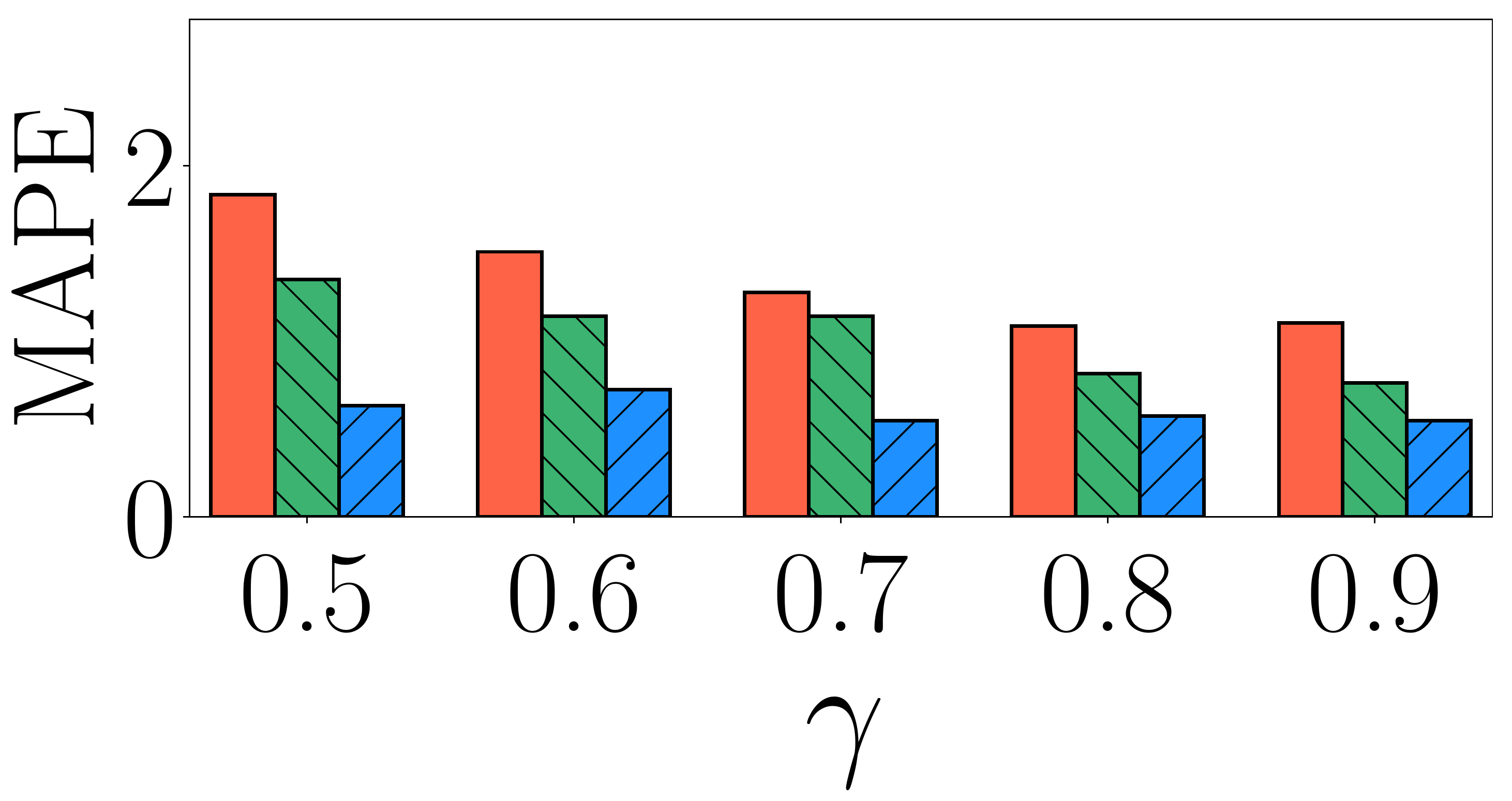} \label{fig:yahoo-gamma-err}}
	\subfloat[\bag (error vs. $\gamma$)]{%
		\includegraphics[width=.25\textwidth]{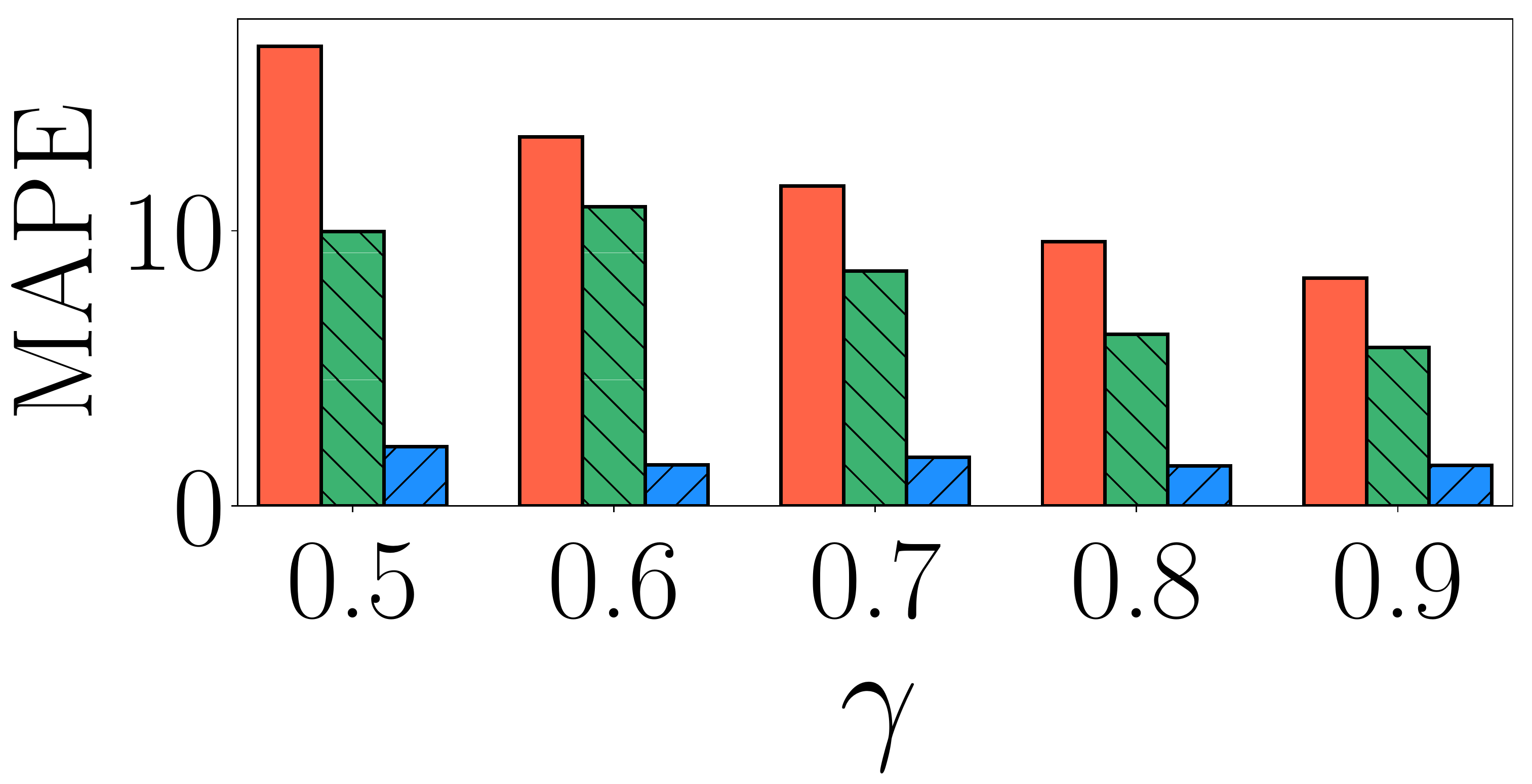} \label{fig:bag-gamma-err}}
	\subfloat[\yahoo (runtime vs. $\gamma$)]{%
		\includegraphics[width=.25\textwidth] {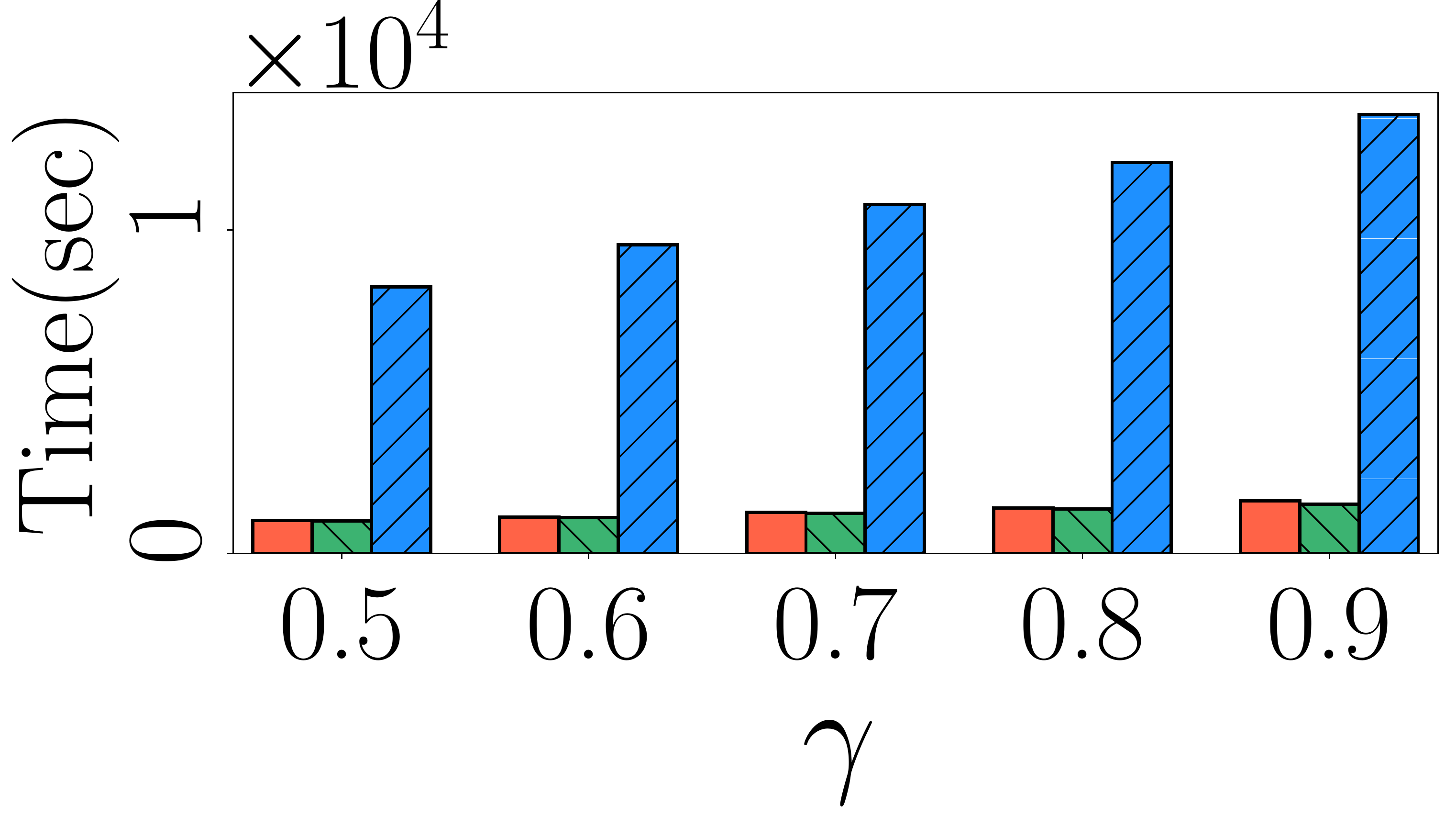} \label{fig:yahoo-gamma-time}}
	\subfloat[\bag (runtime vs. $\gamma$)]{%
		\includegraphics[width=.25\textwidth]{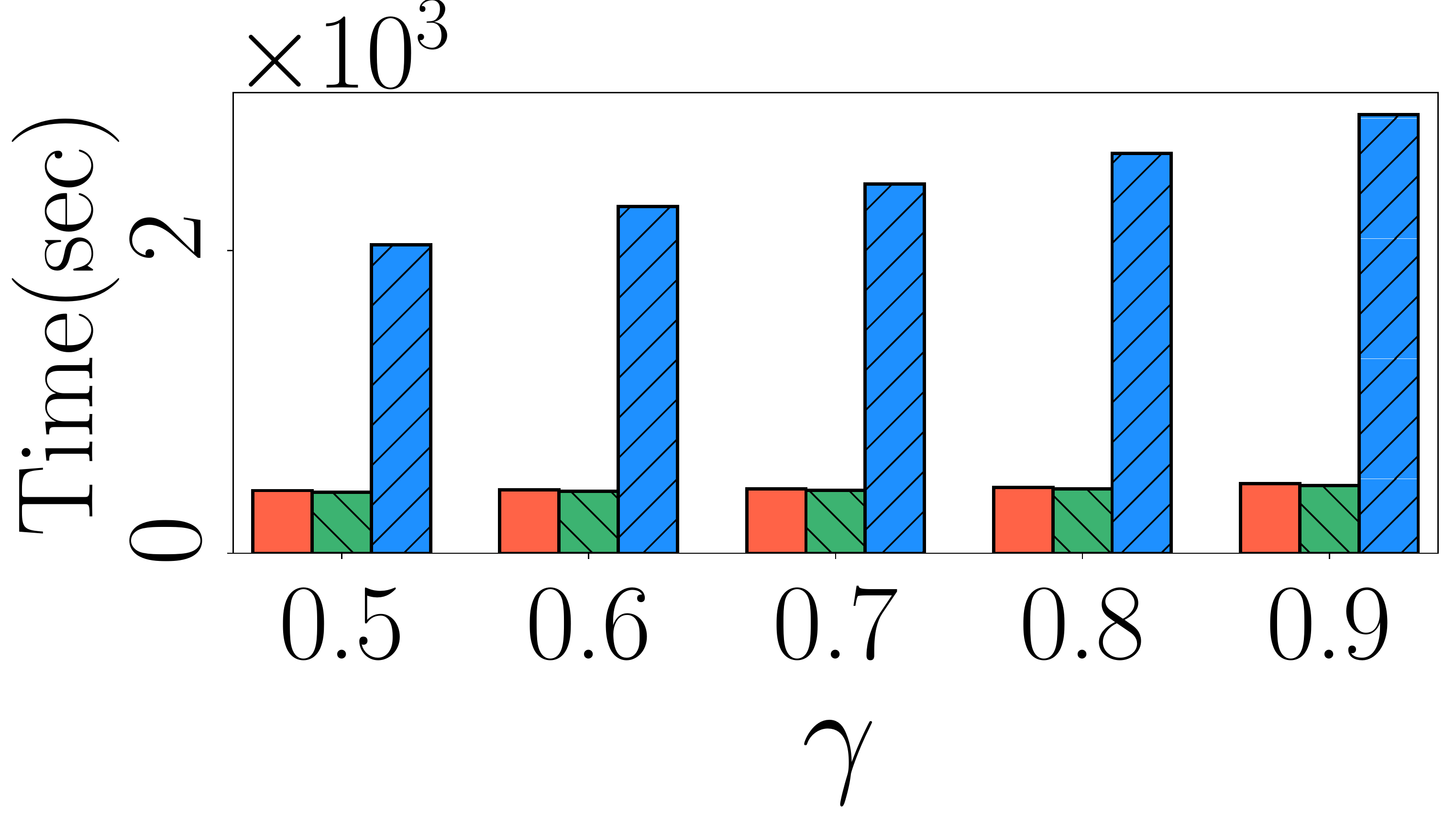} \label{fig:bag-gamma-time}}
	\vspace{-2ex}
	\caption{Accuracy and runtime of \ada, \adasum, and \iada as a function of $\gamma$ where $\ressize = 600K$.}
	\label{fig:gamma-err-time}
	\vspace{-4ex}
\end{figure*}

We experimentally evaluate the infinite window 
and sliding window 
algorithms on real-world temporal bipartite networks with hundreds of millions of edges from a variety of domains, such as social, web, and rating networks. 

{\bf Networks and experimental setup:} 
We used five real-world temporal bipartite networks from the publicly available KONECT repository \cite{K13}, summarized in \cref{tab:networks}.\footnote{http://konect.uni-koblenz.de/} \movie is the ratings by users for movies. \frwiki is a bipartite network of editors and pages of the French Wikipedia where each edge represents an edit. \enwiki is the English version of \frwiki. \yahoo is a ratings by users for songs. \bag is a word-document bipartite network. Note that \bag is not a temporal network, and we generated a stream by randomly permuting the edge set of \bag. \frwiki, \enwiki, and \bag had multiple edges between the same node pairs, and we only considered the first interaction. Edges are read from the stream in the order of timestamps. \cref{fig:bfly-stream} shows the number of butterflies as a function of stream size. 

All streaming algorithms were implemented in \texttt{C++} and compiled with \texttt{g++} compiler using \texttt{-O3} as the optimization level. The source code is publicly available at \cite{srccode}. We run the experiments on a machine equipped with a 2.0 GHz 16-Core Intel E5 2650 processor and 128GB of memory.


\begin{table}[t!]
	\vspace{-3ex}
	\small
	\renewcommand{\tabcolsep}{.5ex}
	\begin{tabular}{|c||r|r|r|r|c|}
		\hline
		Graphs          & \multicolumn{1}{c|}{|E|}     & \multicolumn{1}{c|}{|V| (Left)} & \multicolumn{1}{c|}{|V| (Right)} & \multicolumn{1}{c|}{$\nbfly$} & \multicolumn{1}{c|}{Butterfly density} \\ \hline\hline
		\movie		&	\num{10000054} 	& 	\num{69878} 	& \num{10677} & 1.1T & $1.1 \times 10^{-16}$  \\ \hline
		\frwiki		&	\num{22090703} 	& 	\num{288275} 	& \num{3992426} & 601.2B & $2.5 \times 10^{-18}$  \\  \hline
		\yahoo 		& 	\num{256804235}		&	\num{1000990}	&	\num{624961} & 101.4T & $2.3 \times 10^{-20}$  \\ \hline
		\enwiki		&	\num{122075170} 	& 	\num{3819691} 	& \num{21416395} & 2T & $9.1 \times 10^{-21}$  \\ \hline
		\bag 		& 	\num{483450157}		&	\num{8200000}	&	\num{141043} & 40.8T & $7.4 \times 10^{-22}$ \\ \hline
	\end{tabular}%
	\caption{Properties of the bipartite graphs. $|E|$ is the number of edges, $\nbfly$ the total number of butterflies, and the butterfly density is the ratio $\nbfly/{|E|}^4$.}\label{tab:networks}
\vspace{-1.em}
\end{table}

\subsection{Accuracy\label{sec:exp-accuracy}}
If the true value of the butterfly count is $x > 0$, then the relative error of an estimate $\hat{x}$ is defined as $|x-\hat{x}| / x$ and is usually shown as percent error. We also used MAPE (Mean Average Percentage Error) to measure the accuracy over the entire stream, defined as the average of the relative error, taken over the entire stream.

\cref{fig:err-reservoir} shows the accuracy on the entire stream vs the reservoir size. Larger reservoirs yield better accuracies, as expected. \iada can keep the estimation error around $1\%$ for all networks by storing only $600$K edges in the reservoir. This corresponds to $6\%$, $2.7\%$, $0.49\%$, $0.23\%$, and $0.12\%$ of the total stream sizes for \movie, \frwiki, \enwiki, \yahoo, and \bag, respectively. When the reservoir size is $300$K, \iada yields $3\%$ error for \enwiki and \bag and less than $1\%$ for other networks. As expected \adasum has better accuracy than \ada, and \iada  has the best accuracy. 

\cref{fig:err-stream} shows the relative error at different points in the stream, for a fixed reservoir size. As the stream size increases, the error of \ada and \adasum increase slightly. This can be attributed to the fact that the edge sampling probability $p$ is proportional to $1/t$, where $t$ is the number of edges, and from \cref{lem:ada-bern}, the probability of a given relative error decreases with $p^4 \nbfly\att$. Unless $\nbfly\att$ increases as the fourth power of $t$, the probability of a given relative error increases with the stream size.

{\bf Butterfly Density:} We note the errors of \ada and \adasum for a given reservoir size are roughly correlated with the butterfly density ($\nbfly\att/t^4$ where $t$ is the number of edges). One reason is as follows. Following  \cref{lem:ada-bern}, the probability of a high relative error decreases with $p^4 \nbfly\att$. Setting $p \approx M/t$ where $M$ is the reservoir size, this is $M^4$ times the butterfly density $\frac{\nbfly\att}{t^4}$, showing that the error probability decreases quickly as the butterfly density increases. When the networks are ordered according to increasing butterfly density, we get the order \bag, \enwiki, \yahoo, \frwiki, and \movie. We note that for the same reservoir size, this is exactly the increasing order of accuracy (decreasing order or error) for algorithms \ada and \adasum (\cref{fig:err-stream}). The trend is not so clear for algorithm \iada, since its accuracy depends heavily on the temporal order of the edges within a butterfly. 


\subsection{Runtime and Throughput}
The better accuracy of \iada comes at the cost of increased runtime. From \cref{fig:throughput-reservoir}, we see \iada has the lowest throughput (number of edges processed per second), while \ada and \adasum have similar throughputs. The reason is there is one per-edge butterfly computation for each arriving edge in \iada, where as there is one such computation only for each sampled edge in \ada and \adasum. The throughput decreases as the reservoir size increases, due to the increased cost of per-edge butterfly counting on the reservoir. \iada is able to achieve quite a high throughput, e.g. $6.2 \times 10^5$ edges per second on graph \bag with reservoir size $150$K, making it suitable for practical scenarios.
\adasum always has a slightly higher throughput than \ada. 
Overall, \iada has the best accuracy with a good throughput, while \adasum trades a lower accuracy for a higher throughput.


\subsection{Impact of $\gamma$ on runtime and accuracy}
\label{sec:gamma}
\cref{fig:yahoo-gamma-err,fig:bag-gamma-err} show the accuracy as a function of $\gamma$. As $\gamma$ increases, the average size of the reservoir increases, while the frequency of sub-sampling also increases. The accuracies of \ada and \adasum improve slightly as $\gamma$ increases from $0.5$ to $0.9$ e.g. for graph \yahoo. In contrast, from \cref{fig:yahoo-gamma-time,fig:bag-gamma-time}, the runtime increases for all estimators as $\gamma$ increases.  A value of $\gamma$ of about 0.7 seems to be a good ``middle ground'' since it achieves nearly the best throughput as well as accuracy. 


\begin{table}[!t]
	\vspace{-3ex}
	\small
	\renewcommand{\tabcolsep}{.5ex}
	\resizebox{0.4\textwidth}{!}{%
		\begin{tabular}{|c||r|r|r|r|r|}
			\hline
			Graphs   & \ada & \adasum   & \iada & \mar\cite{Ahmed17}   & \chakra\cite{BC17} \\ \hline\hline
			\movie  	& 1.03                       & 0.72                      & \textbf{0.69}                     & 89.32                     & 103.23                     \\ \hline
			\frwiki 	& 4.37                       & 1.7                       & \textbf{1.68}                     & 63.92                     & 102.36                     \\ \hline
			\yahoo      & 13.97                      & 5.62                      & \textbf{0.78}                     & 43.1                      & 104.48                     \\ \hline
			\enwiki 	& 19.43                      & 9.94                      & \textbf{2.95}                     & 46.65                     & 85.38                      \\ \hline
			\bag        & 103.74 					 & 91.59 					 & \textbf{5.65} 					 & 14.22 					 & 113.77 \\\hline
		\end{tabular}%
	}
	\caption{MAPE (Mean Absolute Percent Error) of different algorithms for $\gamma = 0.8$ and $\ressize = \num{150}$K} 
	\label{tbl:comparison-MAPE}
	\vspace{-3ex}
\end{table}

\subsection{Comparison with prior work}
In this section, we present a comparison between our methods and prior works, including \cite{Ahmed17,Manjunath11,Kane12,BC17}. 

\cref{tbl:comparison-MAPE} presents a comparison with methods: Graph Priority Sampling (\mar)\cite{Ahmed17} and the work of Bera and Chakrabarti (\chakra) (Algorithm 1 from Section 3.1 of \cite{BC17}) for a reservoir size of $150K$ (we found similar results for other reservoir sizes, ranging from $75$K to $600K$). \mar is a subgraph counting algorithm based on a weighted sample of edges, which we specialized for the case of butterfly estimation. We observe that \iada significantly outperforms \mar on all networks, and \ada and \adasum outperform \mar on all networks except \bag. Since \mar stores additional  information for each edge (\textit{weight} and \textit{rank} -- see \cite{Ahmed17} for details), we stored $75K$ edges in the reservoir for \mar to keep its memory equal to our algorithms. The results were quite similar even if we gave twice the memory to \mar. If we used a sample of $150K$ edges in \mar, its error ranged from 7.78\% (\bag) to 90\% (\movie) still much worse than our algorithms, especially \iada.

We compared with \chakra while holding memory equal, even though \chakra is a two-pass streaming algorithm, which cannot be modified to work in a single pass, and works under a more powerful computational model than our algorithms. With a reservoir of size $150K$, we could run $75K$ basic estimators of \chakra, each of which maintained a sample of two edges. On all streams, all of our algorithms outperformed \chakra by significant margins. 

We implemented the algorithm of Manjunath et al. \cite{Manjunath11}, which estimates the number of cycles in a stream using sketches based on complex random variables. To the best of our knowledge, we are the first to implement this algorithm, and even the authors of \cite{Manjunath11} have not provided an implementation. The accuracy of \cite{Manjunath11} is very poor -- the error of the estimator is more than $100\%$ for both graph streams \yahoo and \bag, even with a memory of 600K estimators. In addition, we found their algorithm slow and impractical. The reason is that for each arriving edge in a stream, the algorithm needs to update the values of many complex valued sketches, where the number of sketches is as large as the reservoir size. The throughput of \cite{Manjunath11} on both graphs is only $5.9$ edges per second, $\approx$ $21K$ edges in one hour, meaning that it is about ${10}^5$ times slower than algorithm \iada (see throughput of \iada in \cref{fig:bag-throughput-reservoir,fig:yahoo-throughput-reservoir}). The algorithms of \cite{Kane12} are not practical for handling large graph streams, since they follows a similar structure (this has not been implemented either, to the best of our knowledge).


%

\subsection{Sliding Window}
\begin{figure}[!t]
	\vspace{-6ex}
	\captionsetup[subfigure]{justification=centering}
	\centering
	\subfloat[\movie]{%
		\includegraphics[width=.24\textwidth] 
		{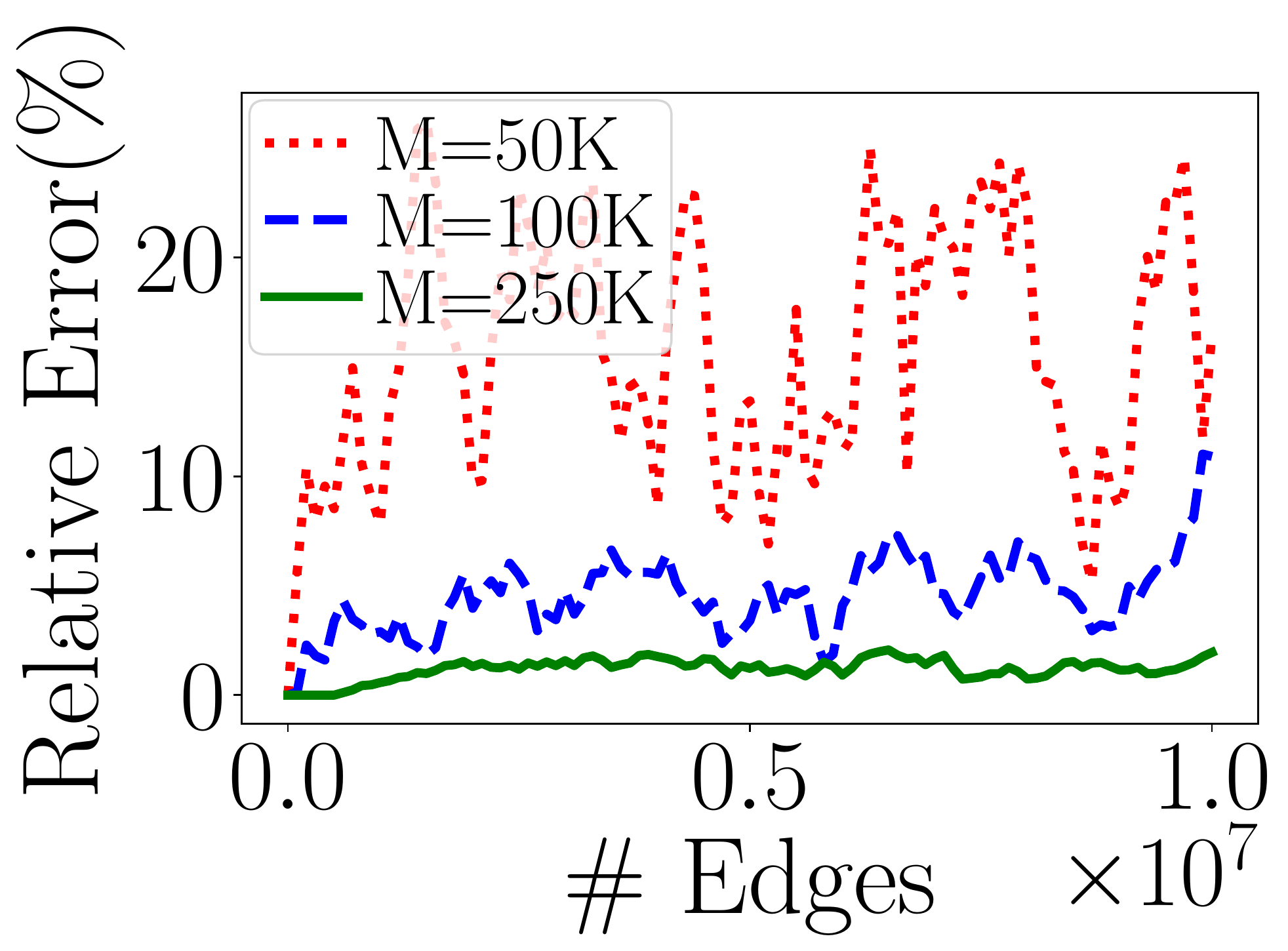} 
		\label{fig:movie-err}}
	\subfloat[\yahoo]{%
		\includegraphics[width=.24\textwidth]
		{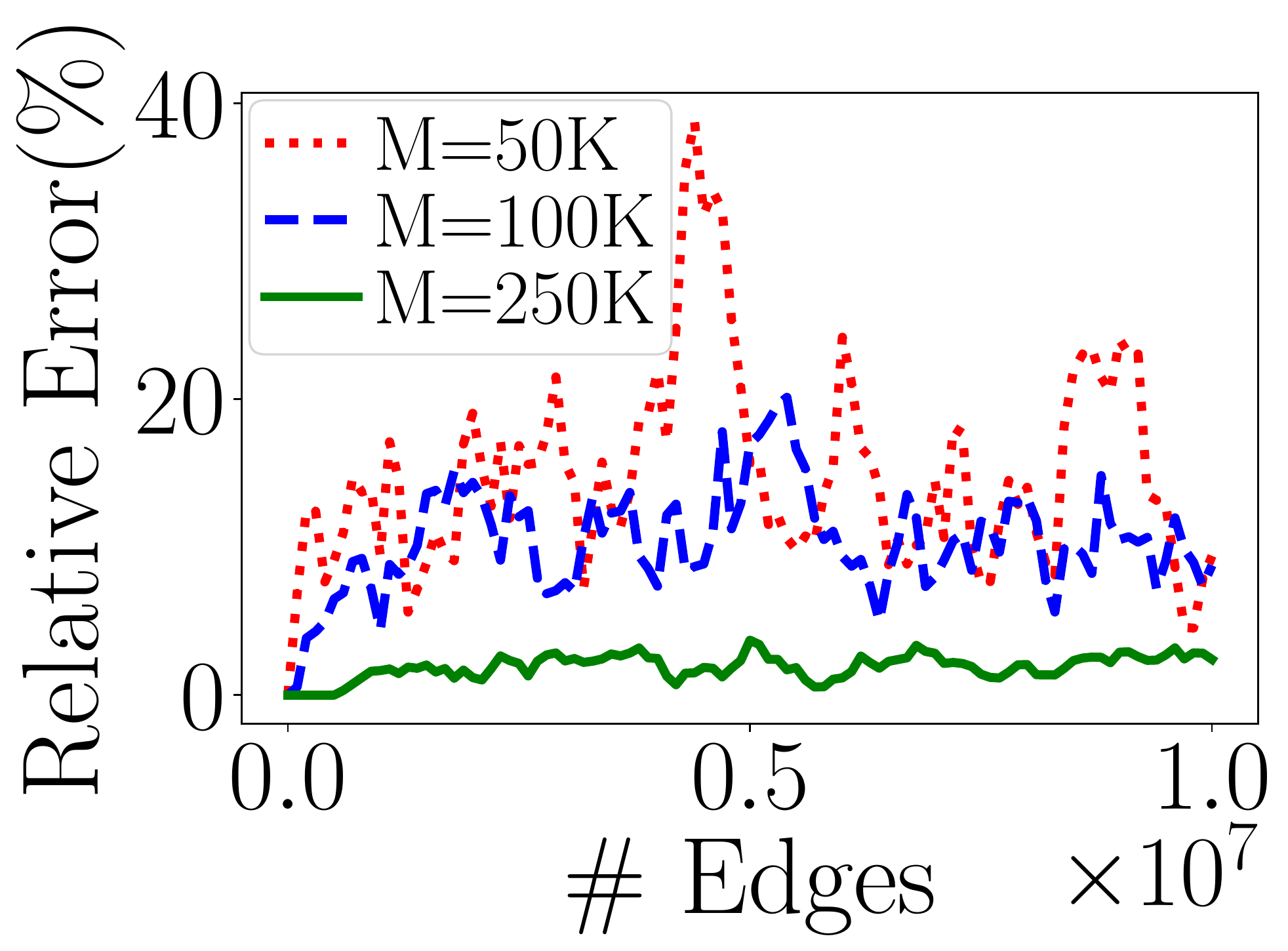} 
		\label{fig:yahoo-err}}
	\vspace{-2ex}
	\caption{Relative error vs. number of edges received for \seqwin. Window size $= 5\times10^{6}$ edges, $\gamma = 0.9$.}
	\label{fig:sld-window-error}
\end{figure}

\cref{fig:sld-window-error} shows the relative error of \seqwin for window size $W = 5$M edges, when the reservoir size is varied from $1\%$ to $5\%$ of $W$. The accuracy improves as the reservoir size increases; when $M$ is $5\%$ of the window size, the relative error is always under $5\%$. The number of butterflies within a window ranges from $5\times10^{10}$ to $10^{11}$ for \movie and $1\times10^{10}$ to $6\times10^{10}$ for \yahoo.  We also experimented with \timewin for time-based windows. We used $30$ queries, and a window size is randomly generated at query time. When the reservoir size $M$ is $10\%$ of the stream size, the average relative error over the queries is $2.55\%$ for movie, $5.52\%$ for \frwiki and $6.32\%$ for \enwiki. This result shows \timewin can achieve good accuracy using memory much smaller than the whole stream. 

\section{Conclusion} 
We presented a lower bound as well as one-pass streaming algorithms for estimating the number of butterflies from a bipartite graph stream. While our lower bound rules out space-efficient algorithms that are accurate on all graph streams, it leaves open the possibility of space-efficient algorithms for graph streams where the number of butterflies is large, such as in every real-world graph stream that we tried. Our algorithms \ada, \adasum, and \iada are based on adaptive random sampling from the graph stream, achieve high accuracy on real-world streams, and are backed by rigorous theoretical guarantees. We also presented algorithms \seqwin and \timewin for sequence-based and time-based sliding windows respectively. This work is one of the first to explore streaming motif counting on bipartite graphs, and leads to many follow-up questions. (1)~Extensions to general motif counting on bipartite graph streams (2)~Can we combine the benefits of improved accuracy as in \iada with the faster runtime of \adasum? (3)~Algorithms for multi-pass and external memory models.

\section{Acknowledgment}
The work of SS, YZ, and ST is supported in part by the National Science Foundation through grants 1527541 and 1725702.

\balance
\bibliographystyle{ACM-Reference-Format}
\bibliography{main}
\end{document}